\documentclass[12pt]{article}

\usepackage{amsmath,amsthm, amssymb,amsfonts,cite}
\usepackage{epsfig,cancel}
\usepackage{dsfont}
\usepackage{mathrsfs}

\numberwithin{equation}{section}

\newcommand{\beq}{\begin{equation}}
\newcommand{\eeq}{\end{equation}}
\newcommand{\ba}{\begin{array}}
\newcommand{\ea}{\end{array}}
\newcommand{\bea}{\begin{eqnarray}}
\newcommand{\eea}{\end{eqnarray}}
\newcommand{\bean}{\begin{eqnarray*}}
\newcommand{\eean}{\end{eqnarray*}}
\newcommand{\eref}[1]{(\ref{#1})}

\newcommand{\comment}[1]{}

\newcommand{\IP}{\mathbb{P}}

\newcommand{\cO}{{\cal O}}
\newcommand{\cN}{{\cal N}}

\newcommand{\cA}{{\cal A}}
\newcommand{\cB}{{\cal B}}
\newcommand{\cC}{{\cal C}}
\newcommand{\cL}{{\cal L}}

\newcommand{\cV}{{\cal V}}

\def\cjn1{{\cA, \cC^*\otimes \wedge^j \cN^*}}
\def\bjn1{{\cA, \cB^*\otimes \wedge^j \cN^*}}
\def\vjn1{{\cA, \cV^*\otimes \wedge^j \cN^*}}
\def\cjn2{{\cA, \cC\otimes \wedge^j \cN^*}}
\def\bjn2{{\cA, \cB\otimes \wedge^j \cN^*}}
\def\vjn2{{\cA, \cV\otimes \wedge^j \cN^*}}

\def\fnote#1#2{\begingroup\def\thefootnote{#1}\footnote{#2}
     \addtocounter{footnote}{-1}\endgroup}

\newtheorem{theorem}{\sf THEOREM}

\newtheorem{lemma}{Lemma}

\newtheorem*{UnLemma}{Lemma}

\begin{document}


\title{{ \bf
Vacuum Varieties, Holomorphic Bundles \\ and Complex Structure Stabilization \\ in Heterotic Theories
}}

\vspace{1cm}

\author{
Lara B. Anderson${}^{1}$,
James Gray${}^{2}$,
Andre Lukas${}^{3}$,
Burt Ovrut${}^{4}$
}

\date{}
\maketitle
\begin{center} {\vskip -0.3cm\small ${}^1${\it Center for the Fundamental Laws of Nature, \\ Jefferson Laboratory, Harvard University, \\ 17 Oxford Street, Cambridge, MA 02138, U.S.A.}\\
  \medskip
    ${}^2${\it Arnold-Sommerfeld-Center for Theoretical Physics, \\
   Department f\"ur Physik, Ludwig-Maximilians-Universit\"at M\"unchen,\\
Theresienstra\ss e 37, 80333 M\"unchen, Germany} \\ 
  \medskip
    ${}^3${\it Rudolf Peierls Centre for Theoretical Physics, Oxford
      University,\\
      $~~~~~$ 1 Keble Road, Oxford, OX1 3NP, U.K.}\\
        \medskip
      ${}^4${\it Department of Physics, University of
      Pennsylvania, \\ Philadelphia, PA 19104-6395, U.S.A.} \\
    \fnote{}{\hskip -0.6cm lara@physics.harvard.edu\\james.gray@physik.uni-muenchen.de\\lukas@physics.ox.ac.uk\\ovrut@elcapitan.hep.upenn.edu} }
\end{center}

\vspace{-1cm}

\abstract{\noindent We discuss the use of gauge fields to stabilize complex structure moduli in Calabi-Yau three-fold compactifications of heterotic string and M-theory. The requirement that the gauge fields in such models preserve supersymmetry leads to a complicated landscape of vacua in complex structure moduli space. We develop methods to systematically map out this multi-branched vacuum space, in a computable and explicit manner. In analysing the resulting vacua, it is found that the associated Calabi-Yau three-folds are sometimes stabilized at a value of complex structure resulting in a singular compactification manifold. We describe how it is possible to resolve these singularities, in some cases, while maintaining computational control over the moduli stabilization mechanism. The discussion is illustrated throughout the paper with explicit worked examples.
 }
\thispagestyle{empty}
\newpage
\setcounter{page}{1}
\tableofcontents

%
%

\section{Introduction}

Since the 1980's great progress has been made in the field of model building utilizing smooth Calabi-Yau compactifications of heterotic string and M-theory. A very large number of models now exist which exhibit standard model charged particle content with no exotics, vector-like or otherwise \cite{Braun:2005ux,Braun:2005bw,Bouchard:2005ag,Anderson:2009mh,Anderson:2011ns,Braun:2011ni,Anderson:2012yf}\footnote{For a selection of work on model building efforts in other heterotic contexts see \cite{Buchmuller:2005jr,Buchmuller:2006ik,Lebedev:2006kn,Kim:2007mt,Lebedev:2007hv,Lebedev:2008un,Nibbelink:2009sp,Blaszczyk:2009in,Blaszczyk:2010db, Kappl:2010yu,Nilles:2012cy,Nibbelink:2012de,Bizet:2013esf, Assel:2009xa,Christodoulides:2011zs,Cleaver:2011ir,Maio:2011qn,GatoRivera:2009yt,GatoRivera:2010xn}. For further work related to the smooth compactifications cited in the text, see \cite{Lukas:1998yy,burt,Braun:2006me,Braun:2005xp,Anderson:2007nc,Anderson:2008uw,Anderson:2010tc,Braun:2013wr}.}. The focus of these phenomenological investigations has now moved on to more detailed questions, such as reproducing the couplings between the particles and the issue of proton stability. 

The field of moduli stabilization has, by contrast, remained more problematic within heterotic theories. The lack of Ramond-Ramond fluxes means that there is naively less structure available to utilize in stabilizing the moduli than in the type II case. This is especially true if one wants to work on a Calabi-Yau three-fold to preserve the model building successes described in the previous paragraph. There are, however, fields beyond the gravitational sector in heterotic compactifications - that is, gauge fields. It has long been known that these could give rise to contributions to the potential for the moduli, even in the case of Calabi-Yau compactifications \cite{Witten:1985bz,Donagi:2009ra}. In the last couple of years it has been realized how to explicitly construct bundles, typically in the hidden sector of the theory, which stabilize complex structure moduli in this context \cite{Anderson:2010mh,Anderson:2011ty}.

It should be stressed that there are moduli associated to the bundles which are introduced to constrain the complex structure in this moduli stabilization mechanism. However, one should not regard this procedure as introducing one set of moduli to stabilize another. Vector bundles (or M5 branes) have to be present in a heterotic compactification to saturate the integrability condition on the Bianchi Identity for the Neveu-Schwarz two-form. As such one, is not introducing new moduli into the problem but, rather, making use of the structure which is already necessarily present. The observation is simply that the combination of the bundle moduli and the complex structure gives an over-counting of the massless fields which are actually present in the compactification. Work is currently being pursued to build these successes into a complete moduli stabilization scenario in the heterotic case, where the bundle moduli and other fields are also stabilized \cite{Anderson:2011cza}.

The mathematics which underpins complex structure stabilization by gauge bundles is due to Atiyah \cite{atiyah}. Atiyah described the following. If one starts with a complex manifold with a given value for its complex structure moduli, and a holomorphic bundle over that space, one can ask what will happen as those fields are varied. In particular, is it possible for the connection on the vector bundle to adjust such that the bundle remains holomorphic no matter how the complex structure moduli are varied? The answer is in the negative. There can be some infinitesimal changes in complex structure which are such that a given holomorphic bundle simply can not adjust to stay holomorphic. Since holomorphy is a requirement for supersymmetry in a heterotic compactification, such a change in the complex structure moduli will break supersymmetry. The resulting potential in the four-dimensional effective theory then constrains the complex structure moduli in that direction - not allowing them to vary without a price in potential energy.

This mechanism for complex structure stabilization works well, and several explicit examples have been elucidated. There is, however, a practical problem in applying the formalism discussed by Atiyah to the physical problem of moduli stabilization. Within combined complex structure and bundle moduli field space, there is typically a complicated structure of vacua to the potential induced by the holomorphy of a given bundle. In physics we are interested in this entire system of vacua. We are especially interested in any vacua which restrict all of the complex structure - which are point like in that moduli space. The problem is that the formalism developed by Atiyah, and described in \cite{Anderson:2010mh,Anderson:2011ty} in this context, does not easily allow us to map out all of this structure.

The Atiyah formalism tells us about fluctuations of moduli fields {\it around a given starting point}. We choose a point in field space, perform a complex calculation, and this tells us which moduli are fixed at that point. If we wanted to map out a complete system of vacua, as described in the previous paragraph, we would have to repeat this calculation for a continuum of points! If we somehow know where in complex structure space to start the analysis, the formalism will confirm whether the moduli are stabilized. But to find that initial point we are reduced to guess work - not a tantalising prospect in a high-dimensional field space.

The goal of this paper is to provide a scanning mechanism which allows us to systematically map out the entire vacuum space carved out by the condition of holomorphy in a large class of example bundles. The techniques we will describe provide a description of the vacuum loci in complex structure moduli space as an algebraic variety. We will illustrate this scanning procedure with several explicit examples and will show that even simple bundles can lead to a rich structure of vacua in the moduli space of the Calabi-Yau compactifications.

Once a description of the vacuum space has been obtained as an algebraic variety, a number of established computational tools are available to study its structure. One of the features which is quickly apparent in studying such systems is that in some, but not all, of the branches of the vacuum space the complex structure values to which one is restricted correspond to Calabi-Yau three-folds which are singular. This leads to a natural question. Can we make sense of these singular branches in the vacuum space? In other words, can we resolve the singularities in the Calabi-Yau three-fold while keeping analytical control of our description of the complex structure stabilization mechanism? The answer to this question is in the affirmative in some cases, and in the final portion of this paper we explain exactly how this is achieved.

The structure of the rest of this paper is as follows. In the next section we describe the procedure which allows us to map out the vacuum structure induced by bundle holomorphy on complex structure moduli space. We begin with a general discussion in subsection 2.1 and give an explicit example in subsection 2.2. In section 3 we describe how to resolve singularities which sometimes appear in the Calabi-Yau in this moduli fixing mechanism. This section culminates in subsection 3.4 where we explicitly resolve the singularities in some of the branches of the vacuum space found in subsection 2.2, maintaining our control over the moduli stabilization mechanism as we do so. We conclude and discuss future work in section 4, while a series of technical appendices introduce some mathematical results which are required in the main text of the paper.

\section{Vacuum Varieties and Bundle Holomorphy}\label{vac_var_sec}

We want to compute whether or not a vector bundle is holomorphic as a function of complex structure. Our starting point is to focus on classes of bundles where the complex structure dependence can be isolated in a particularly calculable manner.

Line bundles are extremely simple with regard to the complex structure dependence of their holomorphy. To see this, we briefly recall the relevant portions of Atiyah's discussion. Given a vector bundle $V$ over a complex manifold $X$ with tangent bundle $TX$, Atiyah defines a new object ${\cal Q}$ by extension:
\begin{eqnarray} \label{at1}
0 \to V \otimes V^{\vee} \to {\cal Q} \to TX \to 0 \ .
\end{eqnarray}
Writing the long exact sequence associated to the short exact sequence \eqref{at1}, one obtains  
a relationship between the cohomology groups of $V$, $TX$ and ${\cal Q}$. Assuming that $X$ is a Calabi-Yau manifold and, hence, that $H^{0}(X,TX)=H^{3}(X,TX)=0$, the long exact sequence is given by
%
\begin{eqnarray} \label{at2}
0 \to H^1(X,V\otimes V^{\vee}) \to H^1(X, {\cal Q}) \to H^1(X, TX) {\to} H^2(X, V \otimes V^{\vee}) {\to} \dots 
\end{eqnarray}
%
In physical langauge, Atiyahs states that the unstabilized complex structure moduli and vector bundle moduli are now combined into a single cohomology, $H^1(X,{\cal Q})$. 
Extracting the relevant information from \eqref{at2}, we see that 
\begin{equation}
H^1(X,{\cal Q}) = H^1(X,V \otimes V^{\vee}) \oplus \ker\big(H^1(X, TX) \stackrel{\alpha}{\to} H^2(X, V \otimes V^{\vee}) \big) \ ,
 \label{at3}
 \end{equation}
 where
 %
%
\begin{equation}
\alpha=\big[F^{1,1} \big] \in H^{1}(V \otimes V^{\vee} \otimes TX^{\vee})
\label{burt2}
\end{equation}
is the cohomology class of the field strength of the connection on the vector bundle $V$, the so-called Atiyah class.

In the case where $V$ is a line bundle on a Calabi-Yau three-fold, $H^2(X, V \otimes V^{\vee})=H^2(X, {\cal O}_{X}) =0$. It follows that the target space of $\alpha$ is empty, ${\rm ker}(\alpha)=H^{1}(X,TX)$ and, hence,  $H^1(X, {\cal Q})$ contains the entirety of $H^1(X, TX)$. That is, no complex structure moduli are stabilized by demanding that a line bundle be holomorphic. In other words, line bundles always remain holomorphic for any complex structure, the connections on them adjusting as necessary as the complex structure moduli change.
Since the holomorphy of line bundles is independent of the complex structure, they have no direct application in the stabilization mechanism we are investigating. However, this very simplicity makes them extremely useful in pursuing our goal of finding classes of bundles where the complex structure dependence can be isolated in a computable manner.

There exist many techniques for building non-Abelian vector bundles out of line bundles using exact sequences. Examples include two term monads and extension bundles, as well as more complicated objects which are often not given specific names. The idea is simply that one writes down a sequence, short exact or otherwise, where all but one of the entries are sums of line bundles. The remaining entry is then taken to be the non-Abelian bundle one is defining. The position of the non-Abelian bundle in the sequence specifies which construction one is using. In such constructions, since the line bundles themselves have no dependence on the complex structure moduli, all of the dependence of the holomorphy of the non-Abelian bundle on those fields can only arise from the maps in the sequences. These maps are given by elements of cohomology classes of sums of line bundles. Thus, we have exchanged the problem of computing bundle holomorphy as a function of complex structure for that of computing cohomology groups of line bundles as a function of those variables. 

There is a large class of examples where such computations are eminently tractable. We will phrase the discussion here in terms of Calabi-Yau three-folds which are complete intersections in products of projective spaces (CICYs). Analogous results can be stated for many other Calabi-Yau constructions, including the hypersurfaces in toric ambient spaces. 
CICYs are defined as the common solution space of a system of polynomial equations in an ambient space given by  ${\cal
  A}=\mathbb{P}^{n_1}\times\dots\times\mathbb{P}^{n_m}$. Let us write the canonically normalized K\"ahler form of each projective space as
$J_{{\mathbb{P}}^{n_{r}}}$. Then line bundles on ${\cal A}$ can be denoted by ${\cal O}_{\cal
  A}(k^1,\ldots ,k^m)={\cal
  O}_{\mathbb{P}^{n_1}}(k^1)\times\ldots\times{\cal
  O}_{\mathbb{P}^{n_m}}(k^m)$, where ${\cal
  O}_{\mathbb{P}^{n_r}}(k^r)$ is the line bundle associated to the
divisor which is Poincar\'e dual to $k^r J_{{\mathbb{P}}^{n_{r}}}$.
We need $K=\sum_{r=1}^mn_r-3$ polynomial equations to define a three-fold as a complete intersection within such an ambient space. Denote the
multi-degrees of these polynomials by ${\bf q}_i=(q^1_i,\ldots
,q^m_i)$, where $q^r_i$ is the degree of the $i$'th polynomial
in the coordinates of the $r$'th projective space. The configuration matrix of a CICY then simply arranges this data in the convenient form
\beq\label{config} \left[\ba{c|cccc}
  \IP^{n_1} & q_{1}^{1} & q_{2}^{1} & \ldots & q_{K}^{1} \\
  \IP^{n_2} & q_{1}^{2} & q_{2}^{2} & \ldots & q_{K}^{2} \\
  \vdots & \vdots & \vdots & \ddots & \vdots \\
  \IP^{n_m} & q_{1}^{m} & q_{2}^{m} & \ldots & q_{K}^{m} \\
  \ea\right]_{m \times K}\; .  \eeq
For such a three-fold to be a Calabi-Yau manifold, the conditions
\beq
\sum_{j=1}^K q^{r}_{j} = n_r + 1 \qquad \forall \; r=1, \ldots, m\;
. \label{c10} 
\eeq 
 must be satisfied. Note that the CICYs defined in this way are simply connected manifolds.

Line bundles on such Calabi-Yau three-folds are defined by their first Chern class. As discussed above, we want to focus on those line bundles whose cohomology can be readily computed as a function of complex structure. As such, we will restrict ourselves to the so-called ``favorable" line bundles. These are defined to be line bundles $\cal{L}$ on $X$ whose first Chern class is a two-form which descends from the ambient space ${\cal A}$. That is,
\begin{eqnarray}
c_1({\cal{L}})= a^r J_r \quad {\rm where}\quad J_{r} = J_{{\mathbb{P}}^{n_{r}}}|_{X}
\end{eqnarray}
for some set of integers $a^r$. We will denote the associated line bundle on the ambient space by ${ \cal L}_{\cal{A}}={\cal O}(a^r)$.
Choosing favorable line bundles means that the complex structure dependence of their cohomology groups can be explicitly calculated using what is called the Koszul sequence. We will see how this is achieved in the next subsection.

To recapitulate the logic, we want to build a non-Abelian bundle $V$ on a CICY $X$ out of favorable line bundles. In doing so, all complex structure dependence of the holomorphy of $V$ will be encoded in the maps of its defining sequence, since the line bundles themselves exhibit trivial dependence on these fields. These maps will be described by elements of line bundle cohomology groups. Thanks to the use of favorable line bundles in the construction, the complex structure dependence of these cohomology groups will be explicitly computable using the Koszul sequence.
The analysis is sufficiently complicated that it is  best illustrated by focussing on one type of bundle construction. As such we will, in the next subsection, carry out the kind of computations we have been describing here for the case of an extension of two favorable line bundles.

Before we move on to the next section we should mention that there are two other important properties that any vector bundle describing a heterotic vacuum should exhibit, besides holomorphy. These are that the bundle should be slope poly-stable and should have a second Chern class compatible with the anomaly cancelation condition of the theory stemming from the integrability condition on the Bianchi Identity for the Neveu-Schwarz threeform field strength. For every explicit example presented in this paper these two conditions are satisfied in at least some sub-cone of the K\"ahler moduli space. We shall not, therefore, discuss this issue further and shall concentrate instead on the holomorphy issues which are central to the discussion of this article.

\subsection{Complex Structure Dependence and Cohomology}

Consider a rank two vector bundle $V$ defined by extension as 
\begin{eqnarray} \label{ext1}
0 \to {\cal{L}} \to V \to {\cal{L}}^{\vee} \to 0 \ .
\end{eqnarray}
The line bundle ${\cal{L}}$ will be assumed to be favorable\footnote{In addition, $\cL$ is chosen to satisfy $\mu(\cL)<0$ where $\mu(\cL)=\int_{X} c_1(\cL)\wedge J \wedge J$ for $J$ a K\"ahler form on $X$. It is straightforward to show that this condition is 1-1 with the slope stability of $V$ for an extension of two line bundles of the form given in \eref{ext1}.}.The extent to which $V$ is not a trivial direct sum is controlled by the extension group $\textnormal{Ext}({\cal{L}}^{\vee}, {\cal{L}})=H^1(X, {\cal{L}}^{\otimes 2})$. Despite the fact that, independently of the choice of complex structure, any line bundle is holomorphic, line bundle cohomology groups such as this one can depend upon complex structure. That is, it is possible for $V$, as defined by certain elements of  $\textnormal{Ext}({\cal{L}}^{\vee}, {\cal{L}})$, to exist as a holomorphic $SU(2)$ bundle only at special loci in complex structure moduli space. 

We are interested, therefore, in computing the cohomology $H^1(X, {\cal{L}}^{\otimes 2})$ as a function of the complex structure moduli of $X$. To simplify notation, and to emphasis that we are computing the cohomology of a line bundle, we define ${\mathscr{L}}= {\cal{L}}^{\otimes 2}$. Note that ${\mathscr{L}}$ remains a favorable line bundle on $X$ which descends from ${\mathscr{L}}_{\cal{A}}$ on ${\cal{A}}$. One can extract the complex structure dependence of the cohomology groups of ${\mathscr{L}}$ by studying the associated Koszul sequence. For $X$ constructed as a co-dimension $K$ complete intersection in some ambient space $\cal{A}$, we have the following exact sequence:
\begin{eqnarray} \label{chappy1}
0 \to \wedge^K {\cal N}^{\vee} \otimes {\mathscr{L}_{\cal{A}}} \to \wedge^{K-1} {\cal N}^{\vee} \otimes {\mathscr{L}_{\cal{A}}} \to \ldots \to {\mathscr{L}_{\cal{A}}} \to {\mathscr{L}} \to 0 \ .
\end{eqnarray}
Here ${\cal N}$ is the normal bundle to $X$ in ${\cal A}$. Sequence \eqref{chappy1} can be split up as
\begin{eqnarray}
0 \to {\cal K} \to {\mathscr{L}_{\cal{A}}} \to {\mathscr{L}} \to 0 \ ,
\end{eqnarray}
where ${\cal K}$ is a bundle defined by the long exact sequence
\begin{eqnarray} \label{mrQ}
0 \to \wedge^K {\cal N}^{\vee} \otimes {\mathscr{L}_{\cal{A}}} \to \wedge^{K-1} {\cal N}^{\vee} \otimes {\mathscr{L}_{\cal{A}}} \to \ldots \to {\cal N}^{\vee} \otimes {\mathscr{L}_{\cal{A}}} \to {\cal K} \to 0 \ .
\end{eqnarray}
The usual analysis of short exact sequences and their associated long exact sequences in cohomology tells us that 
\begin{eqnarray}
\ldots \to H^1({\cal A},{\cal K}) \to H^1({\cal A},{\mathscr{L}_{\cal{A}}}) \to H^1(X,{\mathscr{L}}) \to H^2 ({\cal A},{\cal K}) \to H^2({\cal A},{\mathscr{L}_{\cal{A}}}) \to \ldots  
\end{eqnarray}
This implies that
\begin{eqnarray} \label{coh1}
H^1(X,{\mathscr{L}}) = \textnormal{coker} \left(H^1({\cal A},{\cal K}) \to H^1({\cal A},{\mathscr{L}_{\cal{A}}}) \right) \oplus \textnormal{ker} \left(H^2({\cal A},{\cal K}) \to H^2({\cal A},{\mathscr{L}_{\cal{A}}}) \right) \ .\end{eqnarray}
The general analysis of the complex structure dependence of $H^1(X,{\mathscr{L}})$ from \eqref{coh1} is somewhat involved. Before we describe this in detail, therefore, let us consider a slightly simplified case. This will enable us to see more cleanly how the information of interest is extracted, as well as being sufficiently general to cover the explicit examples given in this paper. We will then return to the general case at the end of this section.

Consider an example with co-dimension one, that is, where $K=1$. The Koszul sequence then becomes the short exact sequence
\begin{eqnarray}
0 \to {\cal N}^{\vee} \otimes {\mathscr{L}_{\cal{A}}} \to {\mathscr{L}_{\cal{A}}} \to {\mathscr{L}} \to 0 \ .
\end{eqnarray}
Hence, ${\cal K}$ is now simply given by ${\cal K}={\cal N}^{\vee} \otimes {\mathscr{L}_{\cal{A}}}$. Expression \eqref{coh1} then becomes 
\begin{eqnarray} 
&& H^1(X,{\mathscr{L}}) = \textnormal{coker} \left(H^1({\cal A},{\cal N}^{\vee} \otimes {\mathscr{L}_{\cal{A}}}) \to H^1({\cal A},{\mathscr{L}_{\cal{A}}}) \right) 
\nonumber \\
&&\qquad\qquad\qquad  \oplus \textnormal{ker} \left(H^2({\cal A},{\cal N}^{\vee} \otimes {\mathscr{L}_{\cal{A}}}) \to H^2({\cal A},{\mathscr{L}_{\cal{A}}}) \right) \ .\label{coh1b}
\end{eqnarray}
We can simplify the discussion further by considering a situation where $H^1({\cal A},\mathscr{L}_{\cal{A}})=0$. In this special case
\begin{eqnarray} \label{simplecoh}
H^1(X,{\mathscr{L}}) = \textnormal{ker} \left(H^2({\cal A},{\cal N}^{\vee} \otimes {\mathscr{L}_{\cal{A}}}) \to H^2({\cal A},{\mathscr{L}_{\cal{A}}}) \right) \ .
\end{eqnarray}

How does one  compute such a cohomology as a function of complex structure? Note that the source and target spaces of the map in \eqref{simplecoh} are {\it ambient space} cohomology groups and, as such, are independent of complex structure. Thus, all of this dependence is found entirely in the map in \eqref{simplecoh} which, from its origin in the Koszul sequence, is determined by the defining relation of the Calabi-Yau three-fold itself.

To describe the cohomology as a function of complex structure, we carry out the following procedure. The source and target ambient space cohomology groups can be described as free polynomial spaces via the theorem of Bott-Borel-Weil (see Appendix \ref{bbw}). More precisely, these cohomology groups can be described in terms of spaces of polynomials in the homogeneous coordinates of the projective factors of the ambient space ${\cal A}$ and also in the inverses of those variables. Explicit examples of this will be given in the next subsection, but for now we continue our discussion in generality. First, construct a general element of the source as a linear combination of the polynomial basis $\{ m^{i} \}$ of $H^2({\cal A},{\cal N}^{\vee} \otimes {\mathscr{L}_{\cal{A}}})$. That is,
\begin{eqnarray} \label{source}
S = s_i m^i \;\; \textnormal{where} \;\; m^i \in H^2({\cal A},{\cal N}^{\vee} \otimes {\mathscr{L}_{\cal{A}}}) \ .
\end{eqnarray}
The $s_i$ are, at present, arbitrary coefficients. Second, form a general defining polynomial for the Calabi-Yau three-fold as a linear combination of the polynomial basis $\{ n^{\alpha} \}$ of $H^0({\cal A},{\cal N})$. That is, 
\begin{eqnarray} \label{mrp}
P = c_{\alpha} n^{\alpha} \;\; \textnormal{where} \;\; n^{\alpha} \in H^0({\cal A},{\cal N}) \ .
\end{eqnarray}
The arbitrary coefficients $c_{\alpha}$ in equation \eqref{mrp} represent a redundant description of the complex structure moduli. $P$ is a polynomial representation of the map in \eqref{simplecoh}, consistent with the Bott-Borel-Weil descriptions of the source and target spaces.

To determine which elements $S$ are in the kernel of the map, multiply $S$ by $P$ and ask that we get $0$ in the target space $H^2({\cal A},{\mathscr{L}_{\cal{A}}})$. Denoting a basis of this cohomology by $p_I \in H^2({\cal A},{\mathscr{L}_{\cal{A}}})$, we then arrive at the following conditions for an element $S$ to be in the kernel:
\begin{eqnarray} \label{var1}
\textnormal{Coefficient}(s_i c_{\alpha} m^i n^{\alpha}, p_I) \cong \lambda^{i \alpha}_I s_i c_{\alpha}=0 \;\; \forall \; I \ .
\end{eqnarray}
Let us describe in a little more detail what we mean by this. Both $S$ and $P$ are polynomials in the homogeneous coordinates of the ambient projective spaces and their inverses (indeed $P$ is a polynomial just in the coordinates themselves). Their product is therefore a sum of terms which are fractions with numerator and denominator both being monomials in the homogeneous coordinates. To implement \eqref{var1} we first, in each term in $SP$, make all cancelations possible between powers of homogeneous coordinates which appear both in numerator and denominator. In the resulting polynomial (in variables and inverses) we then take the coefficient of each basis element $p_I$ of the target space $H^2({\cal A},{\mathscr{L}_{\cal{A}}})$, as computed using the discussion in Appendix \ref{bbw}. This is what is meant by the left hand side of \eqref{var1}. The rest of this equation is just the definition of $\lambda_I$ and, by setting the result to zero, we focus on the kernel of the map $S$. The resulting equation is bilinear in $s_i$ and $c_{\alpha}$. If we now remember that $H^1(X,{\mathscr{L}})=H^1(X,{\cal{L}}^{\otimes 2})$ is the extension group controlling the bundle \eqref{ext1}, we can see that these equations describe a portion of the vacuum space of the system. 

Let us explain the last comment of the previous paragraph further. The coefficients $s_i$ in $S$ are a parameterisation of the possible extensions describing $V$. The coefficients $c_{\alpha}$ in $P$ are a redundant description of the complex structure moduli. The redundancy in this description is well understood, and can be removed if desired. Equations \eqref{var1}, viewed as describing an algebraic variety in combined ($s_i$,$c_{\alpha}$) space, define the loci of points where the given complex structure, $c_{\alpha}$, is compatible with the existence of a holomorphic extension of the type defined by $s_i$ via \eqref{source}. In other words, equations \eqref{var1} describe the vacuum space of the theory in a redundant description of combined complex structure and extension space with the redundancies being explicitly understood. This is everything we need to analyze the vacuum structure of the system.

At this stage, we have a large set of equations describing the vacuum structure of the system as an algebraic variety. We now need to extract information in a useful form. Fortunately, there is a ready-made set of tools available to do so - those of computational algebraic geometry. For example, one may want to know which complex structure moduli can stabilized. That is, which loci in complex structure moduli space can support holomorphic bundles which can not adjust in any way to stay holomorphic when we perturb the system off of those loci?

\begin{figure}[ht]\centering 
\includegraphics[height=8cm,width=13cm, angle=0]{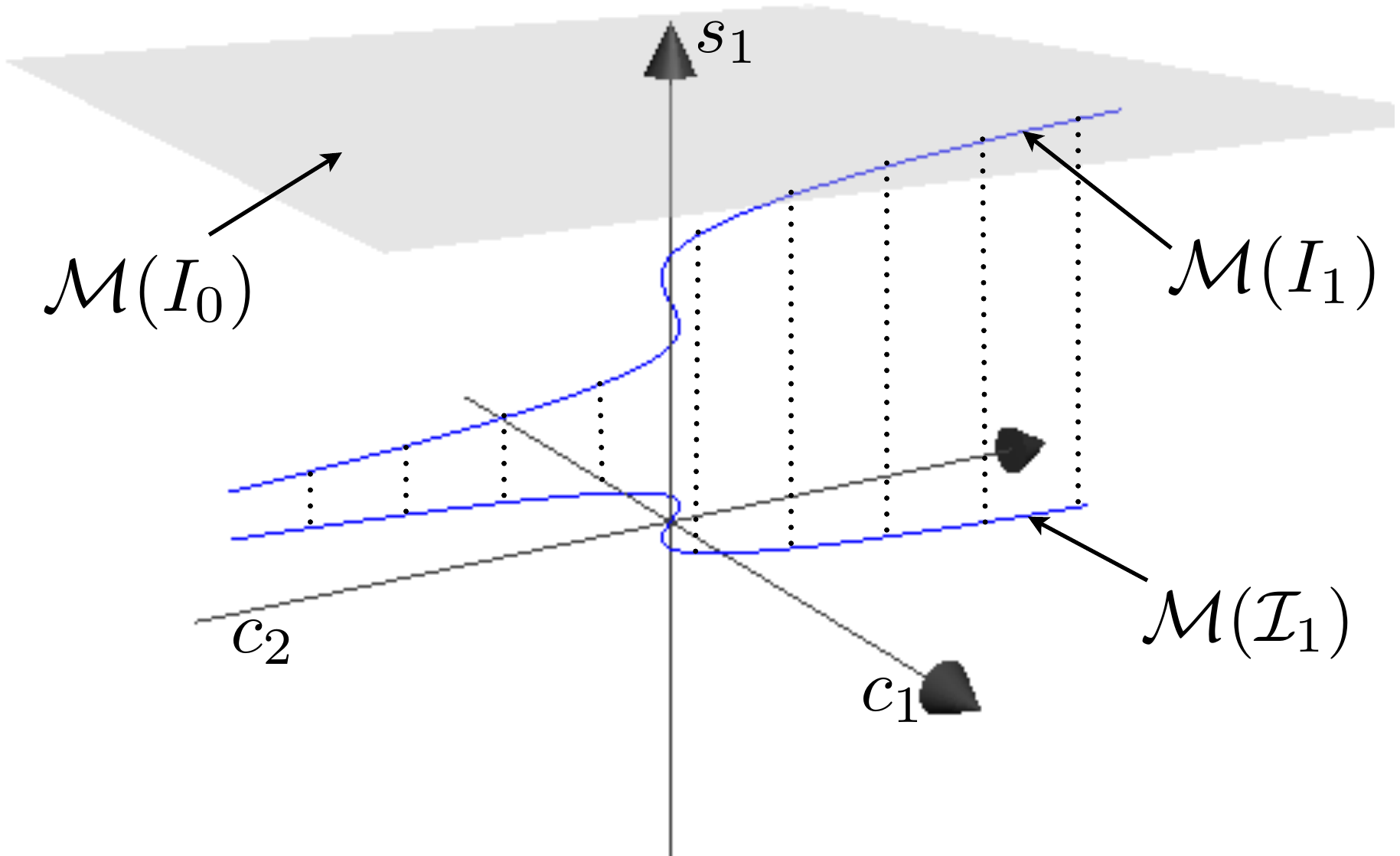}
        \caption{An illustration of the vacuum space}
\label{fig1}
\end{figure}

To perform this analysis, we follow a number of steps.
\begin{itemize}
\item First compute the minimal associated primes of the ideal $\left< \lambda^{i \alpha} s_i c_{\alpha}\right> \subset \mathbb{C}[s_i,c_{\alpha}]$. By definition, the minimal associated primes in this context are a set of ideals, $I_A \subset \mathbb{C}[s_i,c_{\alpha}]$, with $A=1 \ldots N$ where $N$ is the number of irreducible components of the vacuum space. If a given minimal associated prime is generated by a set of polynomials ${\cal G}_A$, that is $I_A=\left< {\cal G}_A \right>$, then the $A$'th irreducible component of the vacuum space is described by the equations obtained by setting each of the polynomials in the set ${\cal G}_A$ to zero. In simple terms, then, this computation takes the one large set of equations \eqref{var1} which describes the entire vacuum manifold and breaks it up into $N$ smaller sets of equations, one describing each irreducible component of that space.

\item By performing a Gr\"obner basis elimination procedure on the $s_i$ variables, one can then find the generators of a new ideal ${\cal I}_A \subset \mathbb{C}[c_{\alpha}]$ where ${\cal I}_A = I_A \cap \mathbb{C}[c_{\alpha}]$. These ideals are generated by the set of equations describing the allowed space of complex structure in each branch of the vacuum manifold of the system. Geometrically, we are taking the varieties describing the irreducible pieces of the vacuum manifold in the combined space of $c_{\alpha}$ and $s_i$ variables and are projecting this down onto the space spanned soley by the complex structure degrees of freedom $c_{\alpha}$\footnote{Strictly speaking, elimination is the algebraic equivalent to the algebraic closure of this projection.}. A graphical depiction of this kind of projection can be found in Figure \ref{fig1}. In this figure, the notation ${\cal M}(J)$, for an ideal $J$, refers to the space of solutions to the equations generating the ideal in the space of its variables.

\item  Once the equations describing each branch of the vacuum space have been found, one can simply count the number of stabilized complex structure moduli by computing the dimension of the associated ideal ${\cal I}_A$. This is achieved by computing a Gr\"obner basis for the ideal, examining the leading monomials, and performing elementary combinatorial manipulations. A detailed discussion of the mathematics of this, and the other computational methods employed in this set of three steps, can be found in this text \cite{decker}.
\end{itemize}

It is important to perform the step of decomposition into associated primes, prior to  projecting onto the subspace of complex structure variables. This is because these two processes do not commute. As a simple example, a solution to \eqref{var1} is always furnished by $s_i=0,\; \forall \; i$. This solution holds for any complex structure. In this branch of the vacuum space, $V$ is simply the Abelian sum of two line bundles and the complex structure is unconstrained. 
By projecting 
this single component onto the subspace of complex structure variables, the full complex structure space is obtained. This situation is depicted in Figure \ref{fig1}. Here ${\cal M}(I_0)$ is the trivial component of the vacuum space just described. We are, of course, interested in the other, more interesting, branches of the vacuum space. Let the blue locus ${\cal M}(I_1)$ denote one less trivial branch of solutions. Projecting ${\cal M}(I_1)$ to the complex structure plane we obtain ${\cal M}({\cal I}_1)$, a restricted locus in moduli space to which we are stabilized. If, however, we project both ${\cal M}(I_0)$ and ${\cal M}(I_1)$ simultaneously, we see that the former will cover the entire complex structure plane and the interesting structure of ${\cal M}({\cal I}_1)$ will be missed. 

The separation in the space of possible extension classes, $s_1$ in Figure \ref{fig1}, is crucial in this discussion. If ${\cal M}(I_1)$ were to lie entirely within ${\cal M}(I_0)$, then physically we would not be able to stabilize the system to the locus ${\cal M}({\cal I}_1)$. Algebraically this subtlety is taken care of by computing minimal associated primes in the first step of the procedure above, and not performing a primary decomposition of a potentially non-radical ideal. This ensures that none of the pieces of the vacuum space which are obtained, ${\cal M}(I_A)$, are wholly embedded within any of the others.

\noindent In the next subsection, we illustrate this procedure with a specific example. We finish this subsection with a further discussion of the general case \eqref{coh1}.

To encompass all possibilities that can arise in a general example, one must extend the above discussion in two important regards. First, we should consider the case where the cokernel in \eqref{coh1b} is non-zero. Second, we should generalize our analysis to higher co-dimension $K$. A non-zero cokernel in \eqref{coh1b} is no more difficult to describe than the kernel discussed above.  The crucial observation is that
\begin{eqnarray} \label{coh2c}
\textnormal{coker}\left(H^1({\cal A},{\cal N}^{\vee} \otimes {\mathscr{L}_{\cal{A}}}) \to H^1({\cal A},{\mathscr{L}_{\cal{A}}}) \right) \cong \textnormal{ker} \left(H^2({\cal A},{\mathscr{L}_{\cal{A}}}^{\vee}) \to H^2({\cal A},{\cal N} \otimes {\mathscr{L}_{\cal{A}}}^{\vee})  \right) \;.
\end{eqnarray}
We can, therefore, express the cokernel in \eqref{coh1b} as a kernel, and compute it in exactly the same manner as we computed the kernel above. With this in mind, expand a general element, $T$, of $H^2({\cal A},{\mathscr{L}_{\cal{A}}}^{\vee})$ in terms of a polynomial basis $\{ q^{a} \}$ as
\begin{eqnarray}
T = t_a q^a \;\; \textnormal{where} \;\; q^a \in H^2({\cal A},{\mathscr{L}_{\cal{A}}}^{\vee}) \;.
\end{eqnarray}
We then repeat the computations that were performed above with $S$. A complete calculation of $H^1(X,{\mathscr{L}})$ then results in a set of equations describing the vacuum space in ($s_i$,$t_a$,$c_{\alpha})$ space. In other words, we obtain an ideal for the vacuum space, $I \subset \mathbb{C}[s_i,t_a,c_{\alpha}]$. As before, this can be decomposed as $I = \cap_A I_A$. An elimination can then be performed to obtain the ${\cal I}_A = I_A \cap \mathbb{C}[c_{\alpha}]$.

The final generalization we need to make is to extend our analysis to the case where the co-dimension of the Calabi-Yau three-fold, $K$, is greater than one. This is straightforward, if laborious. Start with the sequence \eqref{mrQ} defining ${\cal K}$. Inserting appropriate cokernels and kernels, it is possible to break this sequence into a series of short exact sequences. This having been achieved, one can write down the long exact sequences in cohomology associated to each of these short exact sequences. Finally, it is necessary to perform the types of cokernel and kernel computations described in this subsection to obtain an expression for $H^*({\cal A},{\cal K})$. This can then be used as the source spaces in \eqref{coh1} to perform the remainder of the computation.

\subsection{An Example: A Multi-Branch Vacuum Space and its Properties}\label{tetra_branch}

To make the discussion in the proceeding subsection more concrete, let us give an explicit example.  In addition to analyzing the structure of the vacuum space, we will encounter a natural set of questions about these heterotic geometries which will then be addressed in the following section.

We start by presenting the Calabi-Yau three-fold. First consider the tetra-quadric hypersurface in a product of four $\mathbb{P}^1$'s,
\begin{equation}\label{tetrafirst}
\tilde{X}=\left[ \begin{array}{c|c}
{\mathbb P}^1 & 2 \\
{\mathbb P}^1 & 2 \\
{\mathbb P}^1 & 2 \\
{\mathbb P}^1 & 2 
\end{array} \right]^{4,68}_{-124} \ .
\end{equation}
In terms of moduli, $h^{1,1}(\tilde{X})=4$ and $h^{2,1}(\tilde{X})=68$. Let $(x_{i,0}:x_{i,1})$ denote the homogeneous coordinates of the $i$'th $\mathbb{P}^1$. The three-fold $\tilde{X}$ admits a freely acting ${\mathbb Z}_2 \times {\mathbb Z}_4$ symmetry, $\Gamma$, with generators $\gamma_1$ and $\gamma_2$ acting on the ambient space coordinates as 
\begin{eqnarray} \label{symgen}
\gamma_1&:&x_{i,a} \to (-1)^{a+i+1} x_{i,a} \\ \nonumber
\gamma_2&:& x_{i,a} \to x_{\sigma(i),a+i+1}\;\; \textnormal{where} \;\; \sigma=(12)(34) \;.
\end{eqnarray}
Here, we have employed the standard cycle notation in describing the permutations $\sigma$.
One can, therefore, define a smooth Calabi-Yau three-fold  $X=\tilde{X}/\Gamma$. It is this three-fold we will work on. The manifold $X$ is favorable. By this, we mean that the restriction of the K\"ahler forms of the complex projective space factors of the ambient space, ${\cal A}$, to the Calabi-Yau three-fold furnish a basis of harmonic $(1,1)$ forms. Indeed, the manifold $X$ has $h^{1,1}(X)=4$, and $h^{1,2}(X)=10$.

Over $X$, consider the extension bundle
\begin{eqnarray} 
0 \to {\cal O}_{X}(-2,-2,1,1) \to V \to {\cal O}_{X}(2,2,-1,-1) \to 0 \ . \label{egext}
\end{eqnarray}
This is of the form \eqref{ext1}, with the bundle relevant to the associated Ext group being ${\mathscr{L}}={\cal O}_{X}(-4,-4,2,2)$. The Calabi-Yau three-fold $X$ is co-dimension one and, in addition, 
\begin{equation}
H^1({\cal A},{\cal O}_{{\cal{A}}}(-4,-4,2,2))=0 \ . 
\label{burt1}
\end{equation}
Therefore, this is precisely one of the simplified cases considered in the previous subsection. Note that the ambient space line bundle ${\cal O}_{{\cal{A}}}(2,2,-1,-1)$ is equivariant under $\Gamma$ and, thus, it and its dual descend to $X$ intact. Hence, the extension sequence \eqref{egext} is well-defined on $X$.

The normal bundle of $X$ is ${\cal N} = {\cal O}_{{\cal A}}(2,2,2,2)$. Looking at \eqref{simplecoh} and the method outlined in the previous subsection, we see that we must first write down a general element of $H^2({\cal A},{\cal N}^{\vee} \otimes {\mathscr{L}_{\cal{A}}})=H^2({\cal A},{\cal O}_{{\cal{A}}}(-6,-6,0,0))$ as in \eqref{source}. In this case, using Bott-Borel-Weil and remembering that the ambient space for $X$ is $(\mathbb{P}^1 \times \mathbb{P}^1 \times \mathbb{P}^1 \times \mathbb{P}^1)/\Gamma$, we have 
\begin{eqnarray} \nonumber
S&=&s_1 \left( \frac{1}{x_{1,0}^2 x_{1,1}^2 x_{2,0}^2 x_{2,1}^2} \right) + s_3 \left( \frac{1}{x_{1,0}^4 x_{2,0}^2 x_{2,1}^2 } + \frac{1}{x_{1,1}^4 x_{2,0}^2 x_{2,1}^2} + \frac{1}{x_{1,0}^2 x_{1,1}^2 x_{2,0}^4} + \frac{1}{x_{1,0}^2 x_{1,1}^2 x_{2,1}^4}\right) \\ \label{s} &&+ s_2 \left(\frac{1}{x_{1,0}^3 x_{1,1} x_{2,0}^3 x_{2,1}} +\frac{1}{x_{1,0} x_{1,1}^3 x_{2,0}^3 x_{2,1}} + \frac{1}{x_{1,0}^3 x_{1,1} x_{2,0} x_{2,1}^3} +\frac{1}{x_{1,0} x_{1,1}^3 x_{2,0} x_{2,1}}\right) 
\\ \nonumber &&+s_4 \left( \frac{1}{x_{1,0}^4 x_{2,0}^4} +\frac{1}{x_{1,1}^4 x_{2,0}^4} + \frac{1}{x_{1,0}^4 x_{2,1}^4}+\frac{1}{x_{1,1}^4 x_{2,1}^4} \right)\;.
\end{eqnarray}

The defining relation for $X$ is the most general degree $\left[2,2,2,2\right]$ polynomial consistent with the symmetry $\Gamma$ in \eqref{symgen}. This is the polynomial $P$ in \eqref{mrp} for this case, and is explicitly given by the following.
\begin{eqnarray} 
P&=& \text{c}_1 x_{1,0} x_{1,1} x_{2,0} x_{2,1} x_{3,0} x_{3,1} x_{4,0}
   x_{4,1}+   \text{c}_9 \left(x_{1,0}^2 x_{3,0} x_{3,1} x_{4,0} x_{4,1}
   x_{2,0}^2+x_{1,1}^2 x_{3,0} x_{3,1} x_{4,0} x_{4,1} x_{2,0}^2  \right. \nonumber \\
&&  \left.+x_{1,0}^2 x_{2,1}^2
   x_{3,0} x_{3,1} x_{4,0} x_{4,1}+x_{1,1}^2 x_{2,1}^2 x_{3,0} x_{3,1} x_{4,0}
   x_{4,1}\right)+   \text{c}_3 \left(x_{1,1}^2 x_{2,0} x_{2,1} x_{4,0} x_{4,1}
   x_{3,0}^2  \right. \nonumber \\
&&  \left. +x_{1,0} x_{1,1} x_{2,1}^2 x_{3,1} x_{4,0}^2 x_{3,0}+x_{1,0} x_{1,1}
   x_{2,0}^2 x_{3,1} x_{4,1}^2 x_{3,0}+x_{1,0}^2 x_{2,0} x_{2,1} x_{3,1}^2 x_{4,0}
   x_{4,1}\right)+ \nonumber \\
 & &  \text{c}_4 \left(x_{1,0} x_{1,1} x_{2,0}^2 x_{4,0} x_{4,1}
   x_{3,0}^2+x_{1,1}^2 x_{2,0} x_{2,1} x_{3,1} x_{4,0}^2 x_{3,0}+x_{1,0}^2 x_{2,0}
   x_{2,1} x_{3,1} x_{4,1}^2 x_{3,0}  \right. \nonumber \\
&&  \left. +x_{1,0} x_{1,1} x_{2,1}^2 x_{3,1}^2 x_{4,0}
   x_{4,1}\right)+    \text{c}_5 \left(x_{1,0} x_{1,1} x_{2,1}^2 x_{4,0} x_{4,1}
   x_{3,0}^2+x_{1,0}^2 x_{2,0} x_{2,1} x_{3,1} x_{4,0}^2 x_{3,0}  \right. \nonumber \\
&&  \left. +x_{1,1}^2 x_{2,0}
   x_{2,1} x_{3,1} x_{4,1}^2 x_{3,0}+x_{1,0} x_{1,1} x_{2,0}^2 x_{3,1}^2 x_{4,0}
   x_{4,1}\right)+ \text{c}_6 \left(x_{1,0}^2 x_{2,0} x_{2,1} x_{4,0} x_{4,1}
   x_{3,0}^2  \right. \nonumber \\
&&  \left. +x_{1,0} x_{1,1} x_{2,0}^2 x_{3,1} x_{4,0}^2 x_{3,0}+x_{1,0} x_{1,1}
   x_{2,1}^2 x_{3,1} x_{4,1}^2 x_{3,0}+x_{1,1}^2 x_{2,0} x_{2,1} x_{3,1}^2 x_{4,0}
   x_{4,1}\right)+ \nonumber \\
&   &  \text{c}_7 \left(x_{1,1}^2 x_{2,1}^2 x_{3,0}^2 x_{4,0}^2+x_{1,0}^2
   x_{2,1}^2 x_{3,1}^2 x_{4,0}^2+x_{1,1}^2 x_{2,0}^2 x_{3,0}^2 x_{4,1}^2+x_{1,0}^2
   x_{2,0}^2 x_{3,1}^2 x_{4,1}^2\right)+ \nonumber \\
& &  \text{c}_8 \left(x_{1,0}^2 x_{2,1}^2
   x_{3,0}^2 x_{4,0}^2+x_{1,0}^2 x_{2,0}^2 x_{3,1}^2 x_{4,0}^2+x_{1,1}^2 x_{2,1}^2
   x_{3,0}^2 x_{4,1}^2+x_{1,1}^2 x_{2,0}^2 x_{3,1}^2 x_{4,1}^2\right)+ \nonumber \\
  & & \text{c}_2
   \left(x_{1,0} x_{1,1} x_{2,0} x_{2,1} x_{3,0}^2 x_{4,0}^2+x_{1,0} x_{1,1} x_{2,0}
   x_{2,1} x_{3,1}^2 x_{4,0}^2+x_{1,0} x_{1,1} x_{2,0} x_{2,1} x_{3,0}^2
   x_{4,1}^2  \right. \nonumber \\
&&  \left. +x_{1,0} x_{1,1} x_{2,0} x_{2,1} x_{3,1}^2 x_{4,1}^2\right)+   \text{c}_{10}
   \left(x_{1,1}^2 x_{2,0}^2 x_{3,0}^2 x_{4,0}^2+x_{1,1}^2 x_{2,1}^2 x_{3,1}^2
   x_{4,0}^2+x_{1,0}^2 x_{2,0}^2 x_{3,0}^2 x_{4,1}^2  \right. \nonumber \\ \nonumber
&&  \left. +x_{1,0}^2 x_{2,1}^2 x_{3,1}^2
   x_{4,1}^2\right)+
 \text{c}_{11} \left(x_{1,0}^2 x_{2,0}^2 x_{3,0}^2
   x_{4,0}^2+x_{1,1}^2 x_{2,0}^2 x_{3,1}^2 x_{4,0}^2+x_{1,0}^2 x_{2,1}^2 x_{3,0}^2
   x_{4,1}^2 \right. \\  && \left.  +x_{1,1}^2 x_{2,1}^2 x_{3,1}^2 x_{4,1}^2\right)   \label{p}
\end{eqnarray}
Note that, while $X$ has only ten complex structure moduli, there are eleven independent coefficients in \eqref{p}. This is because the $c_{\alpha}$'s are a redundant description of the complex structure. In this case, the only redundancy is that a simultaneous rescaling of all the coefficients in \eqref{p} does not change the locus $P=0$ in $(\mathbb{P}^1 \times \mathbb{P}^1 \times \mathbb{P}^1 \times \mathbb{P}^1)/\Gamma$. Therefore, it does not correspond to a complex structure modulus of $X$.

To construct a general element of $H^1(X,{\mathscr{L}})$ we compute, following the procedure outlined in the previous subsection, the kernel $\textnormal{ker} \left(H^2({\cal A},{\cal N}^{\vee} \otimes {\mathscr{L}_{\cal{A}}}) \to H^2({\cal A},{\mathscr{L}_{\cal{A}}}) \right)$. We multiply \eqref{s} and \eqref{p} together and take the coefficients in the result of a basis of the target cohomology. The target, $H^2({\cal A},{\mathscr{L}_{\cal{A}}})$, is represented by taking combinations of polynomials of degree  $\left[-2,-2,2,2 \right]$ which are invariant under the symmetry action $\Gamma$. This is described in detail in Appendix \ref{bbw}. Since both the source element and the defining relation $P$ respect the symmetry $\Gamma$, their product does as well. As such there is no need to carefully construct a basis for $H^2({\cal A},{\mathscr{L}_{\cal{A}}})$. One can simply take the coefficients of monomials of the correct degree in the product and set these to zero to get the conditions on the $s_{i}$'s such that $S$ is in the kernel of the map. Our procedure then leads to the following equations, the equivalent of \eqref{var1} for this case.
\begin{eqnarray} \nonumber
36s_{2}c_{2}+16s_{1}c_{7}+
96s_{3}c_{8}+96s_{3}c_{10}+
576s_{4}c_{11}=0 \\ \nonumber
36s_{2}c_{2}+96s_{3}c_{7}+
16s_{1}c_{8}+576s_{4}c_{10}+
96s_{3}c_{11}=0 \\ \nonumber
36s_{2}c_{1}+16s_{1}c_{9}+
22s_{3}c_{9}+576s_{4}c_{9}=0 \\ \nonumber
36s_{2}c_{2}+96s_{3}c_{7}+
576s_{4}c_{8}+16s_{1}c_{10}+
96s_{3}c_{11}=0 \\ \nonumber
36s_{2}c_{2}+576s_{4}c_{7}+
96s_{3}c_{8}+96s_{3}c_{10}+
16s_{1}c_{11}=0 \\ \label{fullsys}
16s_{1}c_{3}+36s_{2}c_{4}+
36s_{2}c_{5}+96s_{3}c_{6}=0 \\  \nonumber
36s_{2}c_{3}+16s_{1}c_{4}+
96s_{3}c_{5}+36s_{2}c_{6}=0 \\ \nonumber
36s_{2}c_{3}+96s_{3}c_{4}+
16s_{1}c_{5}+36s_{2}c_{6}=0 \\ \nonumber
96s_{3}c_{3}+36s_{2}c_{4}+
36s_{2}c_{5}+16s_{1}c_{6}=0 \\ \nonumber
16s_{1}c_{2}+36s_{2}c_{7}+
36s_{2}c_{8}+36s_{2}c_{10}+
36s_{2}c_{11}=0 \\ \nonumber
16s_{1}c_{1}+144s_{2}c_{9} =0 
\end{eqnarray}
The equation system in \eqref{fullsys} describes the complete, reducible, vacuum space structure in combined complex structure/bundle moduli space induced by the presence of the bundle \eqref{egext}. We want to further analyze this set of equations to a) see which loci in complex structure moduli space the bundle ${V}$ can stabilize us to and b) what properties the Calabi-Yau three-fold exhibits on these loci.

Our first task is to perform a full primary decomposition (more accurately, we need only compute the minimal associated primes) of the ideal generated by \eqref{fullsys}. This will give us a number  of sets of equations, with each system describing one irreducible component of the vacuum space. Such a primary decomposition can be performed, using the algorithm of Gianni, Trager and Zacharias \cite{GTZ} as implemented in the computer algebra system Singular \cite{singular}. We also make use of the Mathematica interface Stringvacua for this calculation \cite{Gray:2008zs}.
Primary decomposing the ideal generated by \eqref{fullsys}, we find 25 non-trivial branches to the vacuum space. The trivial branches are given by $s_i=0, \;\forall i$ and $c_{\alpha}=0, \;\forall \alpha$ respectively. Clearly, the second of these is not physically relevant. To find the loci in complex structure moduli space that one is restricted to by the 25 interesting branches to the vacuum space, we perform a Gr\"obner basis elimination on each one in turn, as described in the previous subsection. The results are presented in Table \ref{tab1}.

\begin{table}
\begin{eqnarray} \nonumber
\begin{array}{|c|c|c|}
\hline
\textnormal{Equations} & \textnormal{Dimension} & \textnormal{Sing. Dim.} \\
\hline
c_3 - c_4 - c_5 + c_6=c_2 - c_7 - c_8 - c_{10} - c_{11}= c_1 - 4 c_9=0 & 7 & 0 \\ \hline 
c_3+c_4+c_5+c_6=c_2+c_7+c_8+c_{10}+c_{11}=c_1+4 c_9=0 & 7 & 0 \\ \hline
c_9=c_2=c_1=c_7+c_8+c_{10}+c_{11}=c_4+c_5=c_3+c_6=0 & 4 & 0 \\ \hline
c_7-c_8-c_{10}+c_{11}=c_4-c_5=c_3-c_6=c_2=c_1=0 & 5 & 0\\  \hline
c_7-c_8-c_{10}+c_{11}=c_6=c_5=c_4=c_3=c_1 c_8-2 c_2 c_9+c_1c_{10}=0&4&0\\ \hline
c_{11}=c_{10}=c_{9}=c_8=c_7=0 &5 & 0 \\ \hline
c_9=c_6=c_5=c_4=c_3=c_2=c_1=c_8+c_{10}=c_7+c_{11}=0 &1&0\\ \hline
c_9=c_2=c_1=c_8+c_{10}=c_7+c_{11}=c_5+c_6=c_4+c_6=c_3-c_6=0&2&0\\\hline
c_9=c_2=c_1=c_8+c_{10}=c_7+c_{11}=c_5-c_6=c_4-c_6=c_3-c_6=0&2&0\\ \hline
c_{11}=c_{10}=c_9=c_8=c_7=c_2=c_1=c_3-c_4-c_5+c_6=0 &2&0\\ \hline
c_{11}=c_{10}=c_9=c_8=c_7=c_2=c_1=c_3+c_4+c_5+c_6=0 &2&0\\ \hline
c_{11}=c_{10}=c_9=c_8=c_7=c_2=c_1=c_4+c_5=c_3+c_6=0 &1&0 \\ \hline
c_{11}=c_{10}=c_9=c_8=c_7=c_2=c_1=c_4-c_5=c_3-c_6=0 &1&0 \\ \hline
c_{11}=c_{10}=c_9=c_8=c_7=c_2=c_1=c_5+c_6=c_4+c_6=c_3-c_6=0 & 0 & 2\\ \hline
c_{11}=c_{10}=c_9=c_8=c_7=c_2=c_1=c_5-c_6=c_4-c_6=c_3-c_6=0 & 0 & 2\\ \hline
c_{10}-c_{11}=c_8-c_{11}=c_7-c_{11}=c_6=c_5=c_4=c_3=0 &3&0 \\ \hline
c_{10}-c_{11}=c_{8}-c_{11}=c_7-c_{11}=c_6=c_5=c_4=c_3=c_2 c_9 - c_1 c_{11}=0 &2&2 \\ \hline
c_{10}-c_{11}=c_8-c_{11}=c_7-c_{11}=c_4+c_5=c_3+c_6=c_2 c_9-c_1 c_{11}=0 &4&-1 \\ \hline
c_{10}-c_{11}=c_8-c_{11}=c_7-c_{11}=c_5+c_6=c_4+c_6=c_3-c_6=c_2 c_9-c_1 c_{11}=0 & 3 & 1 \\ \hline
c_{10}-c_{11}=c_8-c_{11}=c_7-c_{11}=c_5-c_6=c_4-c_6=c_3-c_6=c_2 c_9-c_1 c_{11}=0 & 3 & 1 \\ \hline
c_8-c_{10}=c_7-c_{11}=c_6=c_5=c_4=c_3=c_2 c_9+50 c_1 c_{10} + 50 c_1 c_{11}=0 & 3 & 0\\ \hline
c_{10}+c_{11}=c_9=c_6=c_5=c_4=c_3=c_2=c_1=c_8+c_{11}=c_7-c_{11}=0 & 0 & 2 \\ \hline
c_{10}+c_{11}=c_9=c_2=c_1=c_8+c_{11}=c_7-c_{11}=c_4-c_5=c_3-c_6=0 &2& 0 \\ \hline
c_{10}+c_{11}=c_9=c_2=c_1=c_8+c_{11}=c_7-c_{11}=c_5+c_6=c_4+c_6=c_3-c_6=0 & 1 & 2 \\ \hline
c_{10}+c_{11}=c_9=c_2=c_1=c_8+c_{11}=c_7-c_{11}=c_5-c_6=c_4-c_6=c_3-c_6=0 & 1 & 2 \\ 
\hline
\end{array}
\end{eqnarray}
\caption{The loci in complex structure moduli space to which the Calabi-Yau three-fold $\tilde{X}/\Gamma$ can be stabilized by the bundle $V$ in equation \eqref{egext}. The ``Dimension'' column refers to the complex dimension of the given locus. The ``Sing. Dim'' column gives the dimension of the singularities in the Calabi-Yau three-fold associated with a generic complex structure in the locus.}
\label{tab1}
\end{table}

The first thing to note about the result in Table \ref{tab1} is that, even in the case of this relatively simple bundle, there are many different loci to which one may be stabilized. It is reasonable to expect the structure to be even richer in more complicated examples. A second important point to make is that, even though some of the loci presented lie inside others in complex structure moduli space, it is still possible to be stabilized to the smaller loci. This is because the extension classes corresponding to these embedded solutions can be different, and the bundle can not undergo a discrete jump in its defining morphisms as it attempts to adjust with complex structure to remain holomorphic.

In addition to the dimension of each locus in complex structure moduli space, Table \ref{tab1} also has an entry labelled ``Sing. Dim.", denoting ``singular dimension''. In restricting the complex structure to lie on a given sub-manifold of moduli space, we specialize the associated Calabi-Yau three-fold. It must be checked whether or not the resulting Calabi-Yau manifold is singular. This can be achieved by forming a ``nodal ideal" on each patch of an open cover of the ambient space ${\cal A}$. This, by definition, is an ideal generated by the defining relations of the three-fold and the $K$ by $K$ minors of the matrix,
\beq \label{H}
H^j_{\mathfrak{m}} = \frac{\partial P^j}{\partial y^{\mathfrak{m}}}  \;.
\eeq
Here the $P^j$, where $j=1 \ldots K$, are the defining relations of the Calabi-Yau three-fold and the $y^{\mathfrak{m}}$, where $\mathfrak{m}=1 \ldots $K$+3$, are the affine coordinates on the coordinate patch of ${\cal A}$ which is being considered. The dimension of this ideal, computed using the same techniques as described in the previous subsection, is the dimension of the singular locus of the Calabi-Yau manifold inside the patch of ${\cal A}$ which is being considered, and in particular a dimension of $-1$ indicates that the three-fold is smooth on this open set. Performing this computation for each patch in an open cover of ${\cal A}$ then maps out the singularity structure of the Calabi-Yau three-fold in detail. Note that it is much more computationally expedient to perform this computation in the patch by patch manner described here than to use the equivalent global formulation that is sometimes employed and which involves just a single dimension computation. This is due to the smaller number of variables involved in the calculation described here providing a large increase in speed given the scaling properties of Gr\"obner basis computations.

Applying this analysis to the case at hand where $K=1$, we find that either the Calabi-Yau associated with a generic complex structure in a locus is singularity-free, or it has a singularity of some dimension\footnote{For any given choice of complex structure there may be singularities of different dimensionalities at various locations on the Calabi-Yau three-fold. We take the dimension of the singularity of the Calabi-Yau manifold for such a complex structure to be that of the largest singular locus on the three-fold.}. Naturally, if we go to special, non-generic points, in the complex structure locus, the Calabi-Yau three-fold will exhibit worse singular behaviour. The set of points where this occurs is measure zero within the locus. The final column in Figure \ref{tab1} gives the minimal dimension of the singular points of a Calabi-Yau three-fold whose complex structure is restricted to each locus\footnote{Note that this is the dimension of sinuglarity induced on the three-fold for generic values of the complex structure on the given locus. For more special points on the complex structure locus, the singularity dimension on the associated Calabi-Yau three-fold may or may not increase.}. 
Only one of the non-trivial, complex structure moduli stabilizing loci in Figure \ref{tab1} corresponds to a smooth Calabi-Yau manifold. This case has the ``Sing. Dim." entry denoted by $-1$. All of the other loci, however, force the Calabi-Yau to become singular. One might imagine that these singular manifolds are physically uninteresting and, hence, can be ignored. However, as we show in detail in the following section, when these singular manifolds can be smoothed out to a singularity free Calabi-Yau three-fold via a ``splitting transition'', it is possible to take our moduli stabilization mechanism through this blowing up process in a consistent manner. Hence, the singular loci in Table \ref{tab1} serve as a ``platform'' for deriving singularity free examples. Clearly, we want to investigate this structure in more detail. This will be the subject of the next section.

We have achieved the goal we set for ourselves at the start of this section. We have mapped out exactly where in complex structure moduli space the vacuum can be stabilized by a large class of vector bundles over Calabi-Yau three-folds. There are many ways in which even more interesting structure can be induced on the complex structure moduli space. For example, having a direct sum of vector bundles in a compactification leads to an intersection of their respective loci as the allowed vacua in complex structure moduli space. At this stage, however, we will content ourselves with the example presented in this section, and will move on to discuss the singularities that can occur in stabilizing complex structure moduli in this manner.

\section{Splitting Transitions and Resolving Singular Points}

\subsection{Resolving Singularities in the Calabi-Yau Threefold}
The previous section makes it clear that the locus in complex structure moduli space to which a holomorphic bundle can restrict a heterotic system can, in some cases, correspond to singular Calabi-Yau three-folds. It is therefore of interest to understand the appearance of these singularities and to decide if they can be resolved.

In general, the question of whether or not singular loci on an arbitrary Calabi-Yau three-fold can be resolved, how to describe this resolved geometry and how the resolution process effects the vector bundle $V$ is a difficult one. In some cases, for example, Calabi-Yau  three-folds defined as hypersurfaces in toric varieties \cite{Kreuzer:2000xy}, the resolution of singularities can be dealt with more or less systematically. For the data set of CICY three-folds in products of projective spaces \cite{Candelas:1987kf} considered in this paper, however, fewer tools are available. However, the resolution of some point-like ``conifold"-type singularities are well understood \cite{Candelas:1989ug}. For these ``splitting transitions''  to a new Calabi-Yau three-fold, $\hat{X}$, where the singularites have been resolved, it will be possible to make substantial progress. In particular, one can straightforwardly identify line bundles on 
$\hat{X}$ whose ``jumping locus" is closely related to that of the original line bundle on $X$. We begin by briefly reviewing the notion of a ``splitting transition" for a CICY three-fold. We then turn our attention to the question of bundles and complex structure stabilization on the geometries related by these transition.

\subsection{A Rapid Review of Splitting Transitions}
Calabi-Yau three-folds defined as complete intersection hypersurfaces in products of projective spaces are known to be simply related to one another via geometric transitions \cite{Candelas:2008wb,Candelas:1989ug,Candelas:2007ac}. Indeed, all 7890 such manifolds \cite{Candelas:1987kf} can be connected through these ``splitting" (conifold) type transitions. Let us briefly recall the basic setup (see \cite{hubsch,Candelas:2007ac} for more detailed reviews). Following the notation of Section \ref{vac_var_sec}, consider the pair of CY three-folds given by
\beq \label{quintsplit}
X_1=[\mathbb{P}^4|~ 5]^{1,101}_{-200}~~~~~,~~~~~
X_2=\left[ \begin{array}{c|c c}
{\mathbb P}^1 & 1 & 1 \\
{\mathbb P}^4 & 1 & 4
\end{array} \right]^{2,86}_{-168} \ .
\eeq
The superscripts on the configuration matrices represent the Hodge numbers $(h^{1,1}, h^{2,1})$ and the subscript denotes the Euler number. Let the homogeneous coordinates of $\mathbb{P}^4$ be $y_0, \ldots y_4$, and those of $\mathbb{P}^1$ be $x_0,x_1$. Then the defining equations of $X_2$ can be written, without loss of generality, as
\bea\label{quinticsplit}
x_0l_1(y)+x_1l_2(y)=0 \\
x_0q_1(y)+x_1q_2(y)=0 \nonumber
\eea
where $l_{1,2}$ and $q_{1,2}$ are linear and quartic polynomials respectively in the homogeneous coordinates of $\mathbb{P}^4$. It is clear that \eref{quinticsplit} can be rewritten as a matrix equation
\begin{eqnarray}
 \left( \begin{array}{cc} l_1 & l_2 \\ q_1 & q_2 \end{array} \right) \left( \begin{array}{c} x_0 \\ x_1 \end{array} \right)=0 \ .
\end{eqnarray}
This has a solution if and only if 
\beq\label{det_var_eg}
{\rm det} \left(\begin{array}{cc} l_1 & l_2 \\ q_1 & q_2 \end{array} \right)=l_1q_2 -l_2 q_1=0 \ .
\eeq
But $l_1q_2 -l_2 q_1$ is nothing less than a quintic polynomial in $\mathbb{P}^4$, exactly the type of polynomial equation defining the quintic three-fold $X_1$. However, $X_1$ and $X_2$ are not the same manifold, precisely because the locus defined by $l_1q_2 -l_2 q_1=0$ is a {\it singular} quintic three-fold\footnote{More precisely, as explained in \cite{Candelas:1989ug}, the quintic three-fold defined by $l_1q_2 -l_2 q_1=0$ is singular at sixteen points. These sixteen points can be blown-up by introducing $\mathbb{P}^1$'s at each of the sixteen nodes. This blowing-up is captured in the configuration matrix by the new $\mathbb{P}^1$ direction in $X_2$. Thus, $X_2$ resolves the singularities of the quintic three-fold in \eref{det_var_eg} and the inclusion of the $16$ $\mathbb{P}^1$'s explains the change in Euler number: $\chi(X_2)=\chi(X_1)+16\chi(\mathbb{P}^1)$.}. The manifolds $X_1$ and $X_2$ are topologically distinct Calabi-Yau three-folds that share a singular locus in their complex structure moduli spaces. Phrased differently, by tuning the complex structure of $X_1$ (that is, shrinking a set of three-cycles, $S^3$, to zero), we arrive at the singular three-fold in \eref{det_var_eg}. This singular geometry can be deformed back to $X_1$ or resolved (by introducing $\mathbb{P}^1$'s at each of the nodes) to give the smooth three-fold $X_2$.

This type of geometric transition relates all of the CICY three-folds to one another. Indeed, it has been speculated \cite{reid_conjecture} that the totality of Calabi-Yau three-folds may be connected by such transitions. For the data set of Calabi-Yau three-folds discussed here, it is important to note that the singular points in moduli space connecting two manifolds may not be just ``conifold" type singularities of the form $f_1f_2 -f_3f_4=0$, as in the case above. Rather, they can be more general ``split" defining equations which correspond to the vanishing loci  of larger $n \times n$ matrices $M$ of polynomials; that is, where $det(M)=0$. These correspond to introducing a new $\mathbb{P}^n$ into the configuration matrix instead of the $\mathbb{P}^1$ in the example above. For instance, another ``split" of the configuration matrix of the quintic $X_1$ is given by
\beq
X_3=\left[ \begin{array}{c|c c c}
{\mathbb P}^2 & 1 & 1 & 1 \\
{\mathbb P}^4 & 1 & 2 & 2
\end{array} \right]^{2,58}_{-112} \ .
\eeq
In this case, the defining equations can be written as
\beq\label{aspliteg}
\left( \begin{array}{c c c}
f^{1}_{1} & f^{1}_{2} & f^{1}_{3} \\
f^{2}_{1} & f^{2}_{2} & f^{2 }_{3} \\
f^{3}_{1} & f^{3}_{2} & f^{3}_{3} 
\end{array} \right)\left( \begin{array}{c}
x_0 \\
x_1 \\
x_2
\end{array} \right)=0 \ ,
\eeq
where $x_{i}$, $i=0,1,2$ are the homogeneous coordinates of $\mathbb{P}^2$ and $f^{1}_{j}$ are general linear functions on $\mathbb{P}^4$ while $f^{2}_{j}$ and $f^{3}_{j}$ are quadratic. The singular locus in complex structure moduli space connecting $X_1$ and $X_3$ is defined by the vanishing of the determinant of the $3 \times 3$ matrix in \eref{aspliteg},
\beq\label{anotherdet}
\textnormal{det} (f^{k}_{j})=0 \ .
\eeq
The ``splitting" process seen in these examples is, in general, not unique. Given a starting CICY, there is a large family of new manifolds that can be constructed in such ways.

To discuss such geometric transitions more generally for the dataset of three-folds at hand, we can define a ``splitting transition" as follows 
\beq\label{gen_split}
X=\left[ {\cal A\,} |\, {\bf c} \; {\cal C} \right] \longrightarrow {\hat X}= \left[ \begin{array}{c|c c c c c}
{\mathbb P}^n & 1 & 1 & \ldots & 1 & {\bf 0} \\
{\cal A} & {\bf c}_1 & {\bf c}_2 & \ldots & \bf{c}_{n+1} & {\cal C} 
\end{array} \right] \; ,\quad {\bf c}=\sum_{{\alpha}=1}^{n+1} {\bf c}_{\alpha}\; .
\eeq
We begin with an initial CICY three-fold, $X$, defined above by a starting configuration matrix of the form $\left[ {\cal A}\, |\, {\bf c}\; {\cal C} \right]$ where ${\cal A}= \mathbb{P}^{n_1} \times \ldots \mathbb{P}^{n_m}$ and ${\bf c}$ and ${\cal C}$ form an $m \times K$ matrix of polynomial degrees for the $K$ equations defining the complete intersection hypersurface. The first column of this matrix, ${\bf c}$, has been explicitly separated from the remainder of the columns, denoted by ${\cal C}$, to facilitate the rest of our discussion.  Since $X$ is a three-fold, $\sum_{r=1}^m n_r -K=3$. We can split $X$ by introducing the new configuration matrix $\hat{X}$ where the vector ${\bf c}$ has been partitioned as the sum of $n+1$ column vectors ${\bf c}_i$ (of dimension $m$) with nonnegative components, as indicated. Since $\hat{X}$ is still a three-fold, the new configuration matrix is $(m+1) \times (K+n)$ dimensional. While the process of going from $X$ to ${\hat X}$ is called ``splitting", the reverse process, in which  ${\hat X} \to X$, is called a ``contraction" \cite{Candelas:2007ac}. In the simple example given in equation \eqref{quintsplit} for example, ${\cal C}$ is an empty matrix, ${\bf c}=5$, ${\bf c}_1=1$ and ${\bf c}_2=4$.

In some cases, a splitting transition of the form \eref{gen_split} will not produce a new Calabi-Yau three-fold, but rather a new description of the same manifold. To see when this is the case we must define the {\it determinental variety} for a general splitting of this type. This definition is taken in direct analogy to \eref{anotherdet} and \eref{det_var_eg}. That is, we write the subset of the defining relations of $\hat{X}$ corresponding to the first $n+1$ columns on the right hand side of \eqref{gen_split} as follows.
\beq\label{gendet}
\left( \begin{array}{cc c c}
f^{1}_{1} & f^{1}_{2} & \ldots & f^{1}_{n+1} \\
f^{2}_{1} & f^{2}_{2} & \ldots & f^{2 }_{n+1} \\
\vdots & \vdots & \ddots & \vdots \\
f^{n+1}_{1} & f^{n+1}_{2}& \ldots & f^{n+1}_{n+1} 
\end{array} \right)\left( \begin{array}{c}
x_0 \\
x_1 \\
\vdots \\
x_{n}
\end{array} \right)=0 \ ,
\eeq
Here $f^{\alpha}_k$ is of degree ${\bf c}_{\alpha}$ for all $k$. The determinental variety is then a special choice of the defining relations for $X$ where the polynomial of degree ${\bf c}$ is taken to be $\textnormal{det}(f^{\alpha}_k)$ and the remaining polynomials, whose degrees are determined by ${\cal C}$, are taken to be arbitrary. We denote the locus in the complex structure moduli space of $X$ where the defining relations take on this determinental form as follows.
\begin{eqnarray}\label{detvardef}
{\cal M}({\cal I}^{det}_{{\hat X}})&=&\{\text{Subset of complex structure moduli space of $X$ such that the first }\nonumber \\
&&\;\;\text{defining equation takes the specialized form}:\textnormal{det}(f^{\alpha}_k)=0 \}\nonumber \subset M^{CS}_{X}\; . 
\end{eqnarray}

If, for all choices of complex structure in ${\cal M}({\cal I}^{det}_{{\hat X}})$, the corresponding three-folds are singular then the splitting transition produces, from $X$, a topologically distinct three-fold $\hat{X}$. In this case,  the ``splitting" of $X$ into ${\hat X}$ is called ``effective". If the determinental variety is smooth for any choice of complex structure in ${\cal M}({\cal I}^{det}_{{\hat X}})$, then $X$ and $\hat{X}$ are diffeomorphic and the splitting is said to be ``ineffective" \cite{Candelas:2007ac}. One simple way to determine whether the splitting is effective is by considering the Euler number. As in the quintic example above, in moving from $X$ to $\hat{X}$ the Euler number changes \cite{Candelas:1989ug} by
\beq
\chi({\hat X})= \chi(X) + 2 (\text{\# of nodes}) \ .
\eeq
Thus, the two manifolds are distinct if and only if the Euler number changes in a splitting transition. In this case, $h^{1,1}(\hat{X})>h^{1,1}(X)$, while $h^{2,1}(\hat{X})< h^{2,1}(X)$.

With these observations, we are ready to use the splitting relationships between the CICY three-folds as a tool to resolve some of the singularities arising from complex structure stabilization in the previous sections. Before we begin, however it is important to give a word of warning on the necessary limitations we will face in comparing Calabi-Yau three-folds related by such transitions. Unlike in Type II string theories \cite{Strominger:1995cz}, dynamical conifold transitions between Calabi-Yau three-folds in heterotic theories are not presently understood in the effective theory. As a result, for all physical vacua, we will stay far away from these singular points in moduli space. The geometric relationships between three-folds will only be used to gain insight into new smooth geometries and to compare independent calculations of complex structure stabilization on both sides of a splitting transition. As a first step towards this goal, we now turn to how divisors, line bundles and their cohomology change under geometric transitions.

\subsection{Divisors, Line Bundles, and Cohomology in Splitting Transitions}

We shall begin this section with the observation that the dimension of the Picard Group increases in a splitting transition as $X \to {\hat X}$. This is to be expected from the fact that, in a conifold-type transition, the Euler number is changing by a positive quantity (that is, $2(\#~\text{of nodes})$) and that three-cycles in $X$ are ``exchanged" for two-cycles in ${\hat X}$. Moreover, recall that in this paper we  consider only ``favorable" divisors and their associated line bundles $\cal{L}$. As defined in Section \ref{vac_var_sec}, these are the restriction of divisors on the ambient space ${\cal A}=\mathbb{P}^{n_{1}}\times \ldots \mathbb{P}^{n_m}$. An inspection of the structure of the Picard groups reveals that favorable divisors on $X$ are ``carried through" the splitting transition and lead to favorable divisors\footnote{Note, however, that since $h^{1,1}({\hat X})> h^{1,1}(X)$, generically the ``new" divisors in ${\hat X}$, that is, those not carried through the transition, will not be favorable with respect to the CICY description of ${\hat X}$.} in ${\hat X}$. 

Recall that we take $J_{r}$, $r=1,\ldots m=h^{1,1}(X)$ to be a basis of harmonic $\{1,1\}$ forms on $X$ and let us label the associated ample divisors by $D_r \subset X$. As was discussed in the previous section, via the line bundle/divisor correspondence, a line bundle on $X$ can be uniquely determined by its first Chern class where ${\cal{L}}=\cO_{X}(a^1 D_1 + \ldots + a^mD_m)=\cO_{X}(a^1,\ldots ,a^m)=\cO_{X}(a^r)$ has $c_1({\cal{L}})=a^r J_r$. The divisors $D_r$ are ``carried through" the splitting transition and can be thought of as also belonging to the  Picard Group of the ``blown-up" geometry ${\hat X}$. In addition, since $h^{1,1}$ increases, there exists a set of $p=h^{1,1}({\hat X})-h^{1,1}(X)$ additional new divisors, $D_{\alpha}$, where $\alpha=\hat{1},\hat{2},\hat{3},\ldots$, associated to the resolution of the singularities in the determinantal variety \eref{detvardef}.

With these definitions in mind, one can now consider how the Chern classes, intersection numbers and line bundle cohomology groups are related as we move from $X$ to a splitting ${\hat X}$. Recall that the triple intersection numbers, $d_{rst}$ on $X$ are defined by
\beq
d_{rst}=\int_{X} J_r \wedge J_s \wedge J_t \ .
\eeq
One useful identity on the triple intersection numbers (see Appendix \ref{split_idens}) is
\beq\label{intersec_result}
d^{X}_{rst}=d^{{\hat X}}_{rst}~ \text{for}~r,s,t=1,\ldots h^{1,1}(X) \ .
\eeq
That is, the intersection numbers of the ``spectator divisors" that are carried through a splitting stay the same across the transition. The remaining triple intersection numbers, $d_{{\alpha}rs}$, $d_{\alpha \alpha r}$, and $d_{\alpha \beta \gamma}$, must be worked out on a case-by-case basis.

Considering a general $\mathbb{P}^n$ splitting as in \eref{gen_split}. Using the notation for line bundles above, one can denote a generic line bundle ${\hat{ \cal{L}}}$ on ${\hat X}$ as ${\hat{ \cal{L}}}=\cO_{{\hat X}}(b^1,\ldots,b^p,a^1, \ldots a^m)$ where $p=h^{1,1}({\hat X})-h^{1,1}(X)=h^{1,1}({\hat X})-m$. For the purposes of this paper, we will be interested in comparing the properties of line bundles that are ``carried through" the splitting transition\footnote{Note that a similar notion of ``carrying" simple bundles through a conifold transition was explored in the mathematics literature in \cite{Chuan:2010si}.}; that is, pairs of line bundles ${\cal{L}}$ and  ${\hat{ \cal{L}}}$ on $X$ and ${\hat X}$, respectively, of the form
\beq\label{linepair}
{\cal{L}}=\cO_X(a^1, \ldots, a^m)~~~~~{\hat{ \cal{L}}}=\cO_{{\hat X}}(0,\ldots 0,a^1, \ldots, a^m) \ .
\eeq
A useful collection of facts regarding such pairs $({\cal{L}},{\hat{ \cal{L}}} )$ can now be compiled. With ${\cal{L}},{\hat{ \cal{L}}}$ defined as in \eref{linepair}, and the triple intersection numbers satisfying  \eref{intersec_result}, it is straightforward to verify that the total Chern character and index of ${\cal{L}}$ and ${\hat{ \cal{L}}}$ satisfy
\begin{align}
&\textnormal{Ch}_{X}({\cal{L}})=\textnormal{Ch}_{{\hat X}}({\hat{ \cal{L}}}) \\
&\textnormal{Ind}_X({\cal{L}})=\textnormal{Ind}_{{\hat X}}({\hat{ \cal{L}}}) \ .
\end{align}
The above equality implies that the coefficients of $\textnormal{Ch}_{{\hat X}}({\hat{ \cal{L}}})$ are identical to those of $\textnormal{Ch}_{X}({\cal{L}})$ when expanded in the basis of ``spectator divisors" $J_r$. Further results (including a useful formula for the relationship between the second Chern class of $X$ and ${\hat X}$) can be found in Appendix \ref{split_idens}.

\subsubsection{Line Bundle Cohomology in Splitting Transitions}
One can, in fact, do better better than an index calculation. For the pair of line bundles $({\cal{L}},{\hat{ \cal{L}}})$, it is possible to compute the exact relation between $H^1(X, {\cal{L}})$ and $H^1({\hat X},{\hat{ \cal{L}}} )$. We state the result in the form of a simple Lemma.
\begin{lemma}\label{lemma1}
Suppose that $X$ and ${\hat X}$ are two Calabi-Yau three-folds realized as complete intersections in products of projective spaces and related by a ``splitting transition'' of the type described in \eref{gen_split}. Let ${\cal{L}}=\cO_{X}(a^{1},\ldots, a^{m})$ be a ``favorable" line bundle on $X$--that is, a line bundle corresponding to a divisor $D \subset X$ such that $D={D}_{\cal{A}}|_X$ is the restriction of a divisor ${D}_{\cal{A}}$ in 
the ambient space. Then the calculation (and dimension) of the cohomology of ${\hat {\cal{L}}}=\cO_{\hat{X}}(0\ldots ,0,a^{1},\ldots, a^{m})$ is identical to that of ${\cal{L}}$ on the ``determinantal locus" (defined by \eref{gen_split} and \eref{detvardef}) shared by the complex structure moduli space of $X$ and ${\hat X}$.
\end{lemma}

\vspace{.5cm}

\noindent The proof of this Lemma is provided in Appendix \ref{app_proof}. Here, we will begin with an illustrative example. Consider the following pair of Calabi-Yau three-folds
\beq\label{cohsplitspace}
X=\left[ \begin{array}{c|c c}
{\mathbb P}^2 & 3 \\
{\mathbb P}^2 & 3
\end{array} \right]^{2,83}_{-162}
~~~~~,~~~~~
{\hat X}=\left[ \begin{array}{c|c c}
{\mathbb P}^1 & 1 & 1 \\
{\mathbb P}^2 & 1 & 2 \\
{\mathbb P}^2 & 2 & 1
\end{array} \right]^{3,51}_{-96} \ .
\eeq
As described in the previous section, these two manifolds share a special (singular) locus in their complex structure moduli spaces. To see this, note that, without loss of generality, the defining equations for ${\hat X}$ can be written as
\bea\label{bicubicdefpol}
z_0 f^{1}_{1}+z_1f^{1}_{2}=0 \\
z_0 f^{2}_{1}+z_1f^{2}_{2}=0 \nonumber
\eea
where $(z_0,z_1)$ are homogeneous coordinates on $\mathbb{P}^1$, $\{ f^{1}_{1}, f^{1}_{2}\}$ are generic polynomials of multi-degree $(1,2)$ in the coordinates of $\mathbb{P}^2 \times \mathbb{P}^2$ and, similarly, $\{f^{2}_{1},f^{2}_{2}\}$ are generic polynomials of degree $(2,1)$. As above, \eref{bicubicdefpol} can be written 
\bea
 \left( \begin{array}{cc} f^{1}_{1} & f^{1}_{2} \\ f^{2}_1 & f^{2}_{2} \end{array} \right) \left( \begin{array}{c} z_0 \\ z_1 \end{array} \right)=0 \ .
\eea
This has a solution if and only if the determinant of the matrix vanishes, giving rise to the special bi-cubic (that is, degree $(3,3)$) hypersurface in $\mathbb{P}^2 \times \mathbb{P}^2$ defined by
\beq\label{split_bicube}
f^{1}_{1}f^{2}_2- f^{1}_{2} f^{2}_{1}=0 \ .
\eeq
As expected from the change in Euler number, the complex structure moduli in this ``determinantal variety" give rise to Calabi-Yau three-folds that are singular at $33$ points and link the complex structure moduli space of $X$ and ${\hat X}$.

Now, having discussed the base geometries, let us consider the favorable line bundle 
\beq
{\cal{L}}=\cO_{X}(-3,3) 
\eeq
defined on $X$.
As outlined in Section \ref{vac_var_sec}, we can describe the cohomology $H^{*}(X,{\cal{L}})$ via a Koszul sequence (see Appendix \ref{coh_appendix} for a review). Using the fact that 
\begin{equation}
{\cal{N}}^{\vee}=\cO_{{\cal A}}(-3,-3) \ ,
\label{lamp1}
\end{equation}
this takes the form
\beq\label{bicubiceg}
0 \to \cO_{{\cal A}}(-3,-3)\otimes {\cal{L}}_{\cal{A}} \to  {\cal{L}}_{{\cal A}} \to {\cal{L}} \to 0 
\eeq
where ${\cal A}={\mathbb P}^{2} \times {\mathbb P}^2$ and ${\cal{L}}_{\cal{A}}=\cO_{{\cal A}}(-3,3)$. Taking the long exact sequence in cohomology associated to \eref{bicubiceg}, we find that $H^0(X,{\cal L})=H^3(X,{\cal L})=0$ and
\begin{align}
&H^1(X, {\cal{L}})=\textnormal{ker}(\phi) \\
&H^2(X, {\cal{L}})=\textnormal{coker}(\phi) 
\end{align}
where
\begin{equation}
\phi: H^2({\cal A}, \cO_{\cal A}(-6,0)) \to H^2({\cal A}, \cO_{\cal A}(-3,3))\label{phidef} \ .
\end{equation}
Here we have used the formalism of Bott-Borel-Weil (see Appendix \ref{coh_appendix}) to establish the vanishings we have stated. We can employ the same formalism to obtain explicit polynomial representatives of elements of source, target and map in \eref{phidef} as
\begin{align}
&H^2({\cal A}, \cO_{\cal A}(-6,0)):~ c_{\left(abc\right)}\frac{1}{x_a x_b x_c} \\
&H^2({\cal A}, \cO_{\cal A}(-3,3)): ~g^{\left(\alpha\beta\gamma\right)}y_{\alpha}y_{\beta}y_{\gamma} \nonumber \\
&\phi \in H^0({\cal A},O(3,3)):~\phi=P_{(3,3)} \nonumber 
\end{align}
where $P_{(3,3)}=0$ is the explicit defining polynomial of the Calabi-Yau hypersurface.
The index $a=0,1,2$ runs over the homogeneous coordinates, $x_a$, of the first ambient $\mathbb{P}^2$ factor and, similarly, $\alpha=0,1,2$ for the coordinates, $y_{\alpha}$, of the second $\mathbb{P}^2$. The calculation to determine the rank of $\phi$ can be performed at any point in the complex structure moduli space of $X$; that is, for any defining polynomial $P_{(3,3)}$. Let us now consider the same calculation for $H^1({\hat X}, {\hat {\cal{L}}})$.

For ${\hat X}$, the Koszul sequence \eref{chappy1} takes the form
\begin{align}
&0 \to \cO_{{\hat {\cal{A}}}}(-2,-3,-3)\otimes \cL_{{\hat {\cal A}}} \to (\cO_{{\hat {\cal A}}}(-1,-1,2)\oplus \cO_{{\hat {\cal A}}}(-1,-2,-1))\otimes \cL_{{\hat {\cal A}}} \to {\cal K}  \to 0 \label{badger1} \\
&0 \to {\cal K} \to \cL_{{\hat {\cal A}}} \to {\hat {\cal{L}}} \to 0 \label{badger2}
\end{align}
where now, ${\hat {\cal A}}=\mathbb{P}^1 \times \mathbb{P}^2 \times \mathbb{P}^2$, and the cokernel ${\cal K}$ has been introduced to split the Koszul sequence into two short exact pieces. Taking the long exact sequences in cohomology associated to \eref{badger1} and \eref{badger2} we get
\begin{align}
& H^1({\hat X},{\hat {\cal{L}}})=\textnormal{ker}({\hat \phi}) \\
& H^2({\hat X},{\hat{\cal{ L}}})=\textnormal{coker}({\hat \phi}) \\
& {\hat \phi}: H^2({\cal {\hat A}}, {\cal K}) \to H^2({\cal {\hat A}},\cO_{{\cal {\hat A}}}(0,-3,3)) \\
& H^2({\hat {\cal A}}, {\cal K}) \simeq H^3({\hat {\cal A}},\cO_{{\cal {\hat A}}}(-2,-6,0)) \label{qdef}
\end{align}
The question now arises, how are $\textnormal{ker}(\phi)$ and $\textnormal{ker}({\hat \phi})$ related? Does the ``jumping" of the $\cL$-valued cohomology on $X$ and the resulting constraints on complex structure tell us anything about the jumping of ${\hat \cL}$ and the complex structure of ${\hat X}$? To answer these questions, the first obstacle one encounters is how to explicitly define the map ${\hat \phi}$. To do this, we turn first to the explicit tensor descriptions of the relevant ambient cohomology groups before looking in detail at the polynomial description. According to Bott-Borel-Weil, one has
\begin{align}
& {\hat \phi}: H^2({\cal {\hat A}}, {\cal K}) \to H^2({\cal {\hat A}},\cO_{{\cal {\hat A}}}(0,-3,3)) \\
&{\hat \phi}: \epsilon_{[AB]}c_{\left(abc\right)} \to g^{\left(\alpha\beta\gamma\right)}\label{epsiloneg}
\end{align}
where $A,B=0,1$ count the homogeneous coordinates, $z_A$, of the $\mathbb{P}^1$ factor appearing in ${\hat \cA}$ and $\epsilon_{[AB]}$ is the two-index fully antisymmetric tensor. Now, as shown in Appendix \ref{coh_appendix}, in this case the map ${\hat \phi}$ takes the form
\beq
f_{1}^{A a (\alpha\beta)}f_{2}^{B (b c) \gamma}
\eeq
where $(f_1,f_2)$ are the two defining polynomials given in \eref{bicubicdefpol} of multi-degree degree $(1,1,2)$ and $(1,2,1)$ respectively. At first sight, one might worry that the antisymmetric tensor appearing in \eref{epsiloneg} might take us outside the realm of simple polynomial descriptions of the cohomology map. Fortunately, however, this is not the case.  Simply writing out the tensorial expression
\beq
c_{\left(abc\right)}(\epsilon_{[AB]}f_{1}^{A a (\alpha\beta)}f_{2}^{B (b c) \gamma})=g^{\left(\alpha\beta\gamma\right)} \ ,
\eeq
it is clear that one can, in fact, still view this as a map between polynomial spaces. Specifically, it is a map from the space of symmetrized ``down index" tensors/inverse-polynomials
\beq
c_{\left(abc\right)}~~\text{equivalently}~~ c_{\left(abc\right)}\frac{1}{x_a x_b x_c}
\eeq 
to symmetrized ``up index'' tensors/ordinary polynomials
\beq
g^{\left(\alpha\beta\gamma\right)}~~\text{equivalently}~~g^{\left(\alpha\beta\gamma\right)}y_{\alpha}y_{\beta}y_{\gamma} \ ,
\eeq
where we have included the antisymmetric tensor in the definition of the map. The map in question, that is, $\epsilon_{[AB]}f_{1}^{A a (\alpha\beta)}f_{2}^{B (b c) \gamma}$, is a very special polynomial indeed. Writing out the contraction with the $\epsilon_{[AB]}$ tensor above, we see that the polynomial map is
\beq\label{split_again}
f_{1}^{1 a (\alpha\beta)}f_{2}^{2 (b c) \gamma}-f_{1}^{2 a (\alpha\beta)}f_{2}^{1 (b c) \gamma}=0  \ .
\eeq
But this is precisely the special bi-cubic hypersurface appearing in the determinantal variety in \eref{split_bicube}! That is, for the calculation of both $H^1(X, \cL)$ and $H^1({\hat X}, {\hat \cL})$ one must compute the kernel of a map from
\beq
\phi~\text{or}~{\hat \phi}: c_{\left(abc\right)}\frac{1}{x_a x_b x_c} \to  g^{\left(\alpha\beta\gamma\right)}y_{\alpha}y_{\beta}y_{\gamma} \ .
\eeq
For $H^1(X, \cL)$, the map in question, $\phi$, is the defining degree $(3,3)$ polynomial of the Calabi-Yau hypersurface $X$ itself. ``Carrying" the line bundle $\cL$ through the splitting transition, we find that $H^1({\hat X}, {\hat \cL})$ is determined by the exact same calculation. However, in this case, {\it the map ${\hat \phi}$ is simply the defining polynomial of the determinantal variety}, the factorized $(3,3)$ polynomial of \eref{split_bicube} and \eref{split_again}. As we will see in the following section, this simple fact will allow us to use our scans/results about holomorphic bundles on $X$ to determine non-trivial information about how holomorphic bundles on a new manifold ${\hat X}$ will restrict its complex structure moduli.

\subsection{An Algorithm for Determining Smooth Stabilized Loci in Complex Structure Moduli Space}\label{alg_sec}

Given these results, we are now in a position to return to the questions raised in Section \ref{vac_var_sec}. Suppose that in using a gauge bundle to fix complex structure moduli, as described in Section \ref{vac_var_sec}, we find that one of the resulting stabilized loci in complex structure moduli space generically gives rise to a singular Calabi-Yau three-fold. We can now use the techniques of the previous section to try to resolve these singularities via a splitting-type transition. But what does the ``jumping" of one line bundle cohomology on $X$ tell us about the jumping of a line bundle cohomology on a ``split" of the starting CICY? Roughly, the idea is as follows. Begin with a generically non-holomorphic vector bundle of the form
\beq
0 \to \cL \to V \to \cL^{\vee} \to 0 \ ,
\eeq
whose defining extension class $Ext^1(\cL^{\vee},\cL)=H^1(X, \mathscr{L})$, where $\mathscr{L}=\cL^{\otimes 2}$, ``jumps" in dimension to non-zero values at the loci ${\cal M}({\cal I})$ in the complex structure moduli space of $X$. Can one induce a complex structure stabilizing bundle ${\hat V}$ on the ``split" manifold ${\hat X}$? A natural starting point is to choose the line bundles in question on ${\hat X}$ to be of the ``spectator" form ${\hat \cL}=\cO_{\hat X}(0,\ldots, 0, a^1,\ldots, a^m)$. Now define the extension
\beq\label{newvhat}
0 \to {\hat \cL} \to {\hat V} \to {\hat \cL}^{\vee} \to 0 \ .
\eeq
with associated extension class $Ext^1(\hat{\cL}^{\vee},\cL)=H^1(X, \hat{\mathscr{L}})$, with $\hat{\mathscr{L}}={\hat{\cL}}^{\otimes 2}$.
Then using Lemma \ref{lemma1} from the previous section, we have at our disposal the following powerful result:
\begin{lemma}\label{lemma2}
Let $(X,\mathscr{L})$ and $({\hat X},{\hat {\mathscr{L}}})$ be defined as in Lemma \ref{lemma1}. The ``jumping" locus $M^{jumping}_{{\hat X}}({\cal I}) \subset M^{cs}_{{\hat X}}$ of the extension class $Ext^{1}_{{\hat X}}=H^1({\hat X},{\hat{\mathscr{ L}}})$ on ${\hat X}$ is given by
\beq\label{intersecloci}
{\cal M}^{jumping}_{{\hat X}}({\cal I}) = {\cal M}({\cal I}^{det}_{{\hat X}}) \cap {\cal M}^{jumping}_{X}({\cal I}) \ ,
\eeq
where ${\cal M}^{jumping}_{X}({\cal I}) \subset M^{cs}_{X}$ is the ``jumping'' locus of $Ext^{1}_{X}=H^1(X,\mathscr{L})$ on $X$.
\end{lemma}

Thus, our approach for finding useful complex structure stabilizing bundles can be outlined algorithmically as follows:
\begin{enumerate}
\item Choose a line bundle $\mathscr{L}$ on $X$, such that $H^1(X,\mathscr{L})=0$ for generic values of the complex structure of $X$.
\item Find all possible sub-loci ${\cal M}({\cal I}_i)$ in the complex structure moduli space of $X$ for which $H^1(X, \mathscr{L})$ ``jumps" to a non-zero value.
\item If all sub-loci lead to smooth Calabi-Yau manifolds, stop. If any sub-loci lead to generically singular three-folds $X$, consider all possible ``splittings" ${\hat X}$ of the form \eref{gen_split}.
\item For each such split manifold ${\hat X}$, consider the induced line bundle ${\hat {\mathscr{L}}}$ as defined in \eref{linepair}.
\item Determine the ``jumping" of ${\hat {\mathscr{L}}}$ on ${\hat X}$ by evaluating the intersection in \eref{intersecloci}.
\item For each ${\hat X}$ which yields a non-trivial intersection in \eref{intersecloci}, check whether the loci ${\cal M}^{jumping}_{\hat X}({\cal I}_A)$ yield smooth Calabi-Yau three-folds. If all are smooth, stop.
\item For each singular ${\cal M}^{jumping}_{\hat X}({\cal I}_B)$ locus on a split ${\hat X}$, iterate the procedure with splittings of ${\hat X}$ in turn.
\end{enumerate}

It should be stressed again that we are not presenting here a complete solution to the problem of singular, stabilized loci in complex structure moduli space raised in Section \ref{vac_var_sec}. Once we have chosen a pair $(X, \mathscr{L})$, and found the loci in complex structure moduli space where $\mathscr{L}$ defines the extension class of a generically non-holomorphic bundle, we do not in general know how to follow the bundle defined by extension through the resolution of singularities of the Calabi-Yau three-fold. Instead, in the algorithm outlined above we provide a prescription for how to analyze a subset of these singular loci where the holomorphic vector bundle can be clearly understood in the resolved geometry. Phrased differently, given a starting bundle and a Calabi-Yau three-fold, the above results give us a list of ``nearby geometries" where we now also have non-trivial information about generically non-holomorphic (complex structure ``stabilizing") vector bundles, whose stabilized loci $M^{jumping}({\cal I})$ may give rise to smooth Calabi-Yau three-folds. 
One benefit of this procedure is that the results of the previous sections, at the very least, save us a great deal of computational effort. Having found a ``complex structure fixing"  bundle $V$ on $X$, we know that such bundles exist on many other three-folds related to $X$ by splitting transitions. The stable loci in the new complex structure moduli spaces can be easily obtained as subspaces of $M^{jumping}_{X}({\cal I})$. 

There are, however, several obstructions that can arise. First, although $V$ on $X$ defined by the extension $0 \to \cL \to V \to \cL^{\vee} \to 0$ leads to the existence of an extension bundle ${\hat V}$ on ${\hat X}$ defined via $0 \to {\hat{\cL}} \to {\hat V} \to {\hat {\cL}}^{\vee} \to 0$, the following non-trivial consistency conditions on a heterotic vacuum must be checked.

\subsubsection{Obstructions to ``Carrying a Bundle Through" a Splitting Transition}\label{obs_sec}
\begin{enumerate}
\item It is possible that for a particular split ${\hat X}$ of $X$, $ {\cal M}({\cal I}^{det}_{{\hat X}}) \cap {\cal M}^{jumping}_{X}({\cal I}) =\emptyset$. In this case, $Ext^1({\hat {\cL}}^{\vee},{\hat \cL})=H^1({\hat X},{\hat {\mathscr{L}}})=0$ everywhere in the complex structure moduli space of ${\hat X}$ and one cannot use the resulting ${\hat V}$ to restrict the complex structure of ${\hat X}$. 
\item The size and structure of the K\"ahler cone changes in moving from $V$ on $X$ to $\hat{V}$ on ${\hat X}$. As a result, the slope stability of ${\hat V}$ is not guaranteed and must be explicitly checked in any given example. Indeed, in some cases ${\hat V}$ may be everywhere unstable on ${\hat X}$ and, hence, not suitable for our purposes.
\item The anomaly cancellation condition $c_2(V) \leq c_2(TX)$ which we impose in our starting geometry does not guarantee that $c_2({\hat V}) \leq c_2(T{\hat X})$. This too must be explicitly checked in each case. Since ${\rm ch}_2({\hat {\mathscr{L}}})_{A}={\rm ch}_2({\hat {\mathscr{L}}})^{rs}d_{rsA}$ for $A=({\alpha, r})$, we have
\bea\label{ch2Liden}
{\rm ch}_2({\hat {\mathscr{L}}})_r={\rm ch}_2(\mathscr{L})_r \\
{\rm ch}_2({\hat{ \mathscr{L}}})_\alpha={\rm ch}_2(\mathscr{L})^{rs}d^{{\hat X}}_{rs\alpha} \ ,
\eea
where we recall that indices $r,s,\ldots $ label the second cohomology classes common to $X$ and $\hat{X}$ while $\alpha,\beta,\ldots $ refer to the new classes which appear on $\hat{X}$. If $V$ satisfies the anomaly cancellation condition, then it is guaranteed that 
\beq\label{anom_cond}
c_2(V)_{r} \leq c_2(TX)_r \Longrightarrow c_2({\hat V})_r \leq c_2(T{\hat X})_r
\eeq
(see Appendix \ref{split_idens}). However, it remains to be checked whether or not
\beq
c_2({\hat V})_\alpha \leq c_2(T{\hat X})_\alpha \ .
\eeq
\item Practically, many of the Gr\"obner basis calculations described in this work are too slow to be completed for large polynomial spaces. As a result, we frequently first quotient the Calabi-Yau three-fold by a discrete symmetry, $\Gamma$, to produce a three-fold, $X/\Gamma$, with fewer parameters. In some case, no compatible discrete automorphism, $\hat{\Gamma}$, exists for the split manifold, ${\hat X}$. As a result, in such cases we lack the computational power to fully execute the algorithm outlined above.
\end{enumerate}

The astute reader may also worry at this point that having begun this exercise with the goal of reducing the number of moduli in the theory, it is dissatisfying to have to introduce new $H^{1,1}$ moduli in moving from $X$ to ${\hat X}$. However, it should be noted that while the splitting transitions do increase the number of K\"ahler moduli, this increase is in general negligibly small compared to the number of complex structure moduli removed. We will see an explicit example of this type in the following section. Having developed the formalism to carry ``complex structure stabilizing" bundles through splitting transitions and the above Algorithm (complete with caveats) in this section, we are ready at last to take another look at the example of Section \ref{vac_var_sec} and its complicated vacuum structure.
 
\subsection{An Example: Resolving Singular Branches in a Multi-Branch Vacuum Space}

In this section, we return to the example given in Section \ref{tetra_branch} and to the loci in complex structure moduli space that were given in Table \ref{tab1}. As noted in Section \ref{vac_var_sec}, many of the loci in Table \ref{tab1} correspond to singular Calabi-Yau three-folds. As an example, in this subsection we will consider one of these singular loci and ask if the associated Calabi-Yau three-fold, which generically has point-like singularities, can be resolved using the algorithmic approach described in Section \ref{alg_sec}?

Consider the fourth locus in Figure \ref{tab1}. This locus in complex structure moduli space is given by the equations
\begin{eqnarray} \label{theloc}
c_1=c_2=0\;,\; c_3=c_6\;,\; c_4=c_5 \;\;,\;\;c_7-c_8-c_{10}+c_{11}=0 \ .
\end{eqnarray}
Following Section \ref{alg_sec}, we must find a collection of three-folds ${\hat X}$ related by splitting transitions to the ``tetra-quadric" of \eref{tetrafirst} and apply the results of Lemma \ref{lemma2}. There are many known splits of the tetra-quadric  in the CICY dataset\footnote{For example, there are $6$ $\mathbb{P}^1$-splits of the tetra-quadric, $10$ $\mathbb{P}^2$-splits and so on.} which we could use in our analysis. However, only one such split ${\hat X}$ is known to be compatible with the chosen symmetry action $\Gamma$ of equation \eqref{symgen}. This is the well-known self-mirror manifold \cite{schoen}, whose determinantal variety relative to the tetra-quadric can be written as 
\begin{eqnarray} \label{splittygoodness}
X= \left[ \begin{array}{c|c} \mathbb{P}^1 & 2 \\ \mathbb{P}^1 & 2 \\ \mathbb{P}^1 & 2 \\ \mathbb{P}^1 & 2 \end{array} \right]^{4,68}_{-124} &\Rightarrow&  P_{(2,2,2,2)} \to f^1_{(2,0,2,0)} f^3_{(0,2,0,2)} - f^2_{(2,0,2,0)} f^4_{(0,2,0,2)} =0  \\ \nonumber
&\Rightarrow& {\hat X} =\left[ \begin{array}{c|cc} \mathbb{P}^1 & 1&1 \\ \mathbb{P}^1 & 2 & 0 \\ \mathbb{P}^1 & 0&2 \\ \mathbb{P}^1 & 2&0 \\ \mathbb{P}^1 & 0 & 2 \end{array} \right]^{19,19}_{0}  \ .
\end{eqnarray}
The manifold ${\hat X}$ has Hodge numbers $h^{1,1}=h^{2,1}=19$ and a the compatible symmetry ${\hat \Gamma}$ generated by
\begin{eqnarray} \label{symgennew}
{\hat \gamma}_1&:&(y_{a} ,x_{i,a}) \to (- y_{a}, (-1)^{a+i+1} x_{i,a})\\ \nonumber
{\hat \gamma}_2&:& (y_{a}, x_{i,a}) \to ((-1)^{a+1} e^{ \pi i /2} y_{a}, x_{\sigma(i),a+i+1})\;\; \textnormal{where} \;\; \sigma=(12)(34)
\end{eqnarray}
where we label the homogeneous coordinates on the new $\mathbb{P}^1$ factor as $y_0$ and $y_1$. In addition, the action of $\hat{\gamma}_1$ inverts the sign of the defining polynomials of ${\hat X}$ and  the action of $\hat{\gamma}_2$ swaps the defining polynomials and inverts the sign of the first. The Hodge numbers ``downstairs" (after quotienting by $\hat{\gamma}$ are $h^{1,1}=h^{2,1}=4$.

We need to find the determinantal locus, ${\cal M}({\cal I}^{det}_{{\hat X}})$, in the complex structure moduli space of the original quotient of the tetra-quadric $X$, where the Calabi-Yau three-fold takes on the determinental form indicated in equation \eqref{splittygoodness}. One begins by writing down a general defining polynomial for $X$, with coefficients $c_{\alpha}$, and equating it to a general defining polynomial of determinental form \eref{detvardef}, respecting symmetry ${\hat \Gamma}$, with coefficients $d_{A}$. Comparing coefficients of monomials in the homogeneous coordinates on both sides, one obtains equations giving the generators of an ideal ${R} \subset \mathbb{C}\left[ d_A, c_{\alpha}\right]$. We then perform a Gr\"obner basis elimination procedure on the $d_A$ variables, similar to those carried out in Section \ref{vac_var_sec}, to form an ideal ${\cal R}=R \cap \mathbb{C} \left[ c_{\alpha} \right]$. The generators of this ideal describe the locus in complex structure moduli space where $X$ takes on the determinental form. In the case at hand, we find 
\begin{eqnarray} \label{detloc}
-c_6^2 c_7 +2 c_3 c_6 c_{10} -c_1 c_{10}^2 -c_3^2 c_{11} + c_1 c_7 c_{11} =0 \\ \nonumber
c_2=c_4=c_5=c_9=0 \;\;,\;\; c_8=c_{10} 
\end{eqnarray}

Using the results of Lemma's \ref{lemma1} and \ref{lemma2}, we now intersect the loci of \eref{intersecloci} to obtain a new ``jumping'' locus on ${\hat X}$. Explicitly, intersecting the locus \eqref{theloc} with the locus where the manifold $X$ takes a determinental form, equation \eqref{detloc}, gives
\begin{eqnarray}\label{thelastone}
c_1=c_2=c_4=c_5=c_9=0\;,\; c_8=c_{10}=c_7-c_{10}+c_{11}\;,\; c_3=c_6 \ .
\end{eqnarray}
We see, therefore, that the algorithm outlined in Section \ref{alg_sec} was successful! Beginning with a $5$-dimensional locus in the complex structure moduli space of $X$ which led to point-like singularities in the Calabi-Yau three-fold, we have obtained a new two-dimensional locus, ${\cal M}^{jumping}_{{\hat X}}({\cal I})$ in \eref{thelastone}, which leads to a completely smooth three-fold ${\hat X}$ in \eref{splittygoodness}. In summary, we have found a new bundle
\beq\label{new_improved_bundle}
0 \to \cO_{\hat X}(0,-2,-2,1,1) \to {\hat V} \to \cO_{\hat X}(0,2,2,-1,-1) \to 0
\eeq
on ${\hat X}$ which fixes that manifold's complex structure to the locus in \eref{thelastone}. This is a resolution of a (singular) stabilized locus \eref{theloc} associated to the bundle $V$ defined in \eref{egext} on $X$. For this locus at least, we have successfully resolved the singularities arising in the complex structure stabilization process. 

Bearing in mind the list of possible obstructions in Section \ref{obs_sec}, it is important to verify that the new bundle, \eref{new_improved_bundle}, on ${\hat X}$ satisfies all the consistency conditions for a good Heterotic compactification. Since we have already discussed the compatible discrete symmetries on $X, {\hat X}$, the only remaining conditions to check (as mentioned in Section \ref{obs_sec}) are anomaly cancellation and slope-stability of ${\hat V}$. Using the fact that $d^{X}_{rst}=d^{\hat X}_{rst}$ for the K\"ahler forms ``carried through" the conifold transition, as well as the identities in \eref{ch2Liden}, for anomaly cancellation it only remains to check
\bea
c_2({\hat V})_{\hat{1}} \leq c_2(T{\hat X})_{\hat{1}}
\eea
For the one new direction (associated to the new $\mathbb{P}^1$ factor in \eref{splittygoodness}) on ${\hat X}/{\hat \Gamma}$. For this example, this inequality is readily verified. Finally, it is straightforward to show (see \cite{Anderson:2009nt,Anderson:2009sw}) that ${\hat V}$ is stable if and only if
\beq
\mu(\hat{\cL})= d^{{\hat X}}_{\hat{1}\hat{1}r}c_1({\hat \cL})^r (t^{\hat{1}})^2 + d^{{\hat X}}_{rst}c_1({\hat \cL})^{r}t^st^t <0
\eeq
for some values of the K\"ahler moduli $(t^{\hat{1}}, t^r)$ in the K\"ahler cone of ${\hat X}$. Once again, this inequality is satisfied for $({\hat X}, {\hat V})$ defined above. Thus, the resolved geometry provides a fully consistent background for a heterotic compactification.

In summary then, for this example we began with $10$ complex structure moduli on $X/\Gamma$ defined by \eref{tetrafirst} and \eref{symgen}. By choosing the generically non-holomorphic bundle $V$ in \eref{egext}, $5$ complex structure moduli were fixed, but leading the resulting Calabi-Yau three-fold to be singular at points. By replacing these singular points with $\mathbb{P}^1$'s, it is possible to arrive at the smooth three-fold ${\hat X}$ in \eref{splittygoodness}, which in the presence of the associated bundle ${\hat V}$ in \eref{new_improved_bundle}, leaves only $2$ complex structure moduli remaining. Thus, finally we have removed $8$ complex structure moduli from the initial problem, though at the cost of introducing one additional K\"ahler modulus in the resolution process. Thus, we have a net moduli reduction of $7$.

Finally, we note that it would be natural to continue in this same vein and explore the resolution of the other 24 singular loci in Table \ref{tab1}. The same split $\hat{X}$ of $X$ resolves several of the loci in the table. However, these all end up stabilizing all but two of the complex structure, exactly the same number as the case described in detail above. This split of the tetra-quadric does not resolve many of the singular loci, even some of those with only point-like singularities in $X$. The three-folds with complex structure restricted to these loci remain singular when taken through the transition. It would be interesting to look for other possible resolving geometries as some of these unresolved loci look promising from the point of view of stabilizing all of the complex structure. For example, the seventh entry in Table \ref{tab1} has only one complex structure left unstabilized and has point-like singularities in $X$. We would therefore only need to fix one extra complex structure degree of freedom in resolving this case to a smooth three-fold. Unfortunately, for the order $8$ symmetry, \eref{symgen}, chosen in this example, ${\hat X}$ in \eref{splittygoodness} is the only split of the tetra-quadric known to preserve this symmetry. Thus, we are unable to perform the same analysis for the remaining loci in Table \ref{tab1}\footnote{It would be satisfying to perform these same stabilizing/splitting analysis ``upstairs" without first quotienting by discrete symmetries, but unfortunately, the Groebner basis calculations involved are too slow to finish with existing algorithms/computer speeds.}. 

\section{Conclusions and Further Work}
Over the past two years it has become clear that the gauge fields in heterotic theories can play an important role for moduli stabilization, particularly for the K\"ahler and complex structure moduli which arise in Calabi-Yau compactifications of the theory. The stabilization of complex structure moduli is related to a complicated web of sub-loci in complex structure moduli space which arises in the presence of gauge fields on the Calabi-Yau manifold. On these ``jumping loci", the bundle moduli space is larger than at generic points in complex structure moduli space. Hence, at such loci, a non-generic choice of bundle obstructs moving off this locus and thereby fixes a certain number of complex structure moduli. 

In this paper, we have focused on two important aspects related to this sub-structure of the moduli space. Firstly, we have presented a computational method to determine the complete web of the jumping loci and have applied this method to a specific example. It turns out that, even for the relatively simple example on the tetra-quadric Calabi-Yau manifold with a rank two vector bundle, the resulting structure is very rich and results in 25 different branches with diverse dimensions. Other examples which can be analyzed with our method show a similarly rich structure and we expect this to be a common feature of complex structure moduli spaces in the presence of gauge bundles. Another, initially unexpected property is that the Calabi-Yau manifold becomes singular on many, although not all of these loci. Of course, the supergravity approximation breaks down in the presence of such singularities and any discussion of moduli stabilization based on such singular loci would be unreliable.

In a second step, we have shown how to deal with this difficulty. It turns out that in many cases of interest, the singularities can be resolved by means of a conifold transition while preserving both the methodology and the results of the original computation. In particular, we have applied this method to one of the branches found for our tetra-quadric example. We were able to blow up the point-like singularities which arise on this branch and determine the corresponding locus on the resolved manifold. More generally, we have a established a universal rule by which the original jumping locus and its counterpart on the split manifold are related by an intersection with the determinental variety. This means that the dimension of the jumping locus always either remains unchanged or decreases under a split, a fact which is of direct relevance for moduli stabilization. 

For the purpose of moduli stabilization, point-like jumping loci are of primary interest since all complex structure moduli can be fixed in this case. Unfortunately, we have not found an explicit example of such a zero-dimensional locus while keeping the Calabi-Yau manifold non-singular, although our tetra-quadric example leads to several singular examples. We are not aware of any in-principle obstruction to the existence of non-singular, point-like loci and expect that they will be found by studying a larger number of examples. Currently, such a larger scale study is limited by the available computer power. 

We should emphasize that, in this paper, we have not attempted to study the dynamics of a conifold transition in the presence of gauge bundles. As is, our method should be interpreted as a way of transferring results for jumping loci from one manifold to another one, related by a conifold transition. However, some of our results may well be useful to clarify the fate of heterotic gauge fields under conifold transitions. We hope to return to this point in a future publication.

\section*{Acknowledgments}
L.~A.~is supported by the Fundamental Laws Initiative of the Center for the Fundamental Laws of Nature, Harvard University. The work of J.~G.~was partially supported by NSF grant CCF-1048082, CiC (SEA-EAGER): A String Cartography. A.~L.~is supported in part by the EC 6th Framework Programme MRTN-CT-2004-503369 and EPSRC network grant EP/I02784X/1.
B.~A.~O. is supported in part by the DOE
under contract No. DE-AC02-76-ER-03071 and the NSF under grant
No. 1001296.

\appendix

\section{Topological Identities in Splitting Transitions}\label{split_idens}

At many points in this text we consider splitting  transitions of the following form (see \eqref{gen_split}).
\beq\label{gen_split_app}
X=\left[ {\cal A}\, |\,{\bf c}\;  {\cal C} \right] \longrightarrow {\hat X}= \left[ \begin{array}{c|c c c c c}
{\mathbb P}^n & 1 & 1 & \ldots & 1 & {\bf 0} \\
{\cal A} & {\bf c}_1 & {\bf c}_2 & \ldots & \bf{c}_{n+1} & {\cal C}
\end{array} \right]\; ,\quad {\bf c}=\sum_{i=1}^{n+1}{\bf c}_i\; .
\eeq

In this short appendix we collect and prove certain identities relating the intersection numbers and second Chern classes of $X$ and $\hat{X}$.

\subsection{Intersection Numbers} 

As in the text, let $J_{r}$, $r=1,\ldots m=h^{1,1}(X)$ be a basis of harmonic $\{1,1\}$ forms on $X$, descending from the K\"ahler forms ${\cal J}_r$ of the ambient space complex projective factors. In addition let ${\cal J}_R$ and $J_R$ be their counterparts for $\hat{X}$, were $R=(\hat{1},r)$ runs over $m+1$ values, the first of which, denoted by $\hat{1}$, referring to the K\"ahler form of $\mathbb{P}^n$ and the remaining values, $r$, to the projective factors in ${\cal A}$, as before.

The intersection numbers of $X$ can be written as follows.
\beq \label{unsplitd1}
d^X_{rst}=\int_{X} J_r \wedge J_s \wedge J_t = \int_{\cal A} ({\cal J}_r \wedge {\cal J}_s \wedge {\cal J}_t) \wedge \mu^X 
\eeq
Here,
\beq \label{unsplitmu}
\mu^X= c^r {\cal J}_r \wedge \mu^{\cal C} \;,
\eeq
where
\beq
\mu^{\cal C} =\wedge^{K}_{a=2} ({\cal C}^{ra}{\cal J}_r) \;.
\eeq

A similar expression holds for $\hat{X}$,
\beq \label{splitd1}
d^{\hat{X}}_{RST}=\int_{X} J_R \wedge J_S \wedge J_T = \int_{\cal A} ({\cal J}_R \wedge {\cal J}_S \wedge {\cal J}_T) \wedge \mu^{\hat{X}} \;,
\eeq
where 
\beq \label{splitmu}
\mu^{\hat{x}} = \wedge_{\Lambda=1}^{n+1}( {\cal J}_{\hat{1}} + c_{\Lambda}^r {\cal J}_r) \wedge \mu^{\cal C} \;.
\eeq

Consider the  intersection numbers $d^{\hat{X}}_{\hat{1}rs}$ on the split manifold. Using expressions \eqref{unsplitd1} and \eqref{unsplitmu}, together with the  integration properties of the ${\cal J}$'s over the respective projective spaces, we find that,
\beq
d^{\hat{X}}_{\hat{1}rs}= \int_{{\cal A}} \mu^{\cal C} \wedge \sum_{\Lambda < \Sigma} (c_{\Lambda}^t {\cal J}_t \wedge c_{\Sigma}^u {\cal J}_u )\wedge {\cal J}_r \wedge {\cal J}_s \;.
\eeq
Defining
\beq \label{dtilde}
\tilde{d}_{uvrs}=\int_{{\cal A}} \mu^{{\cal C}} \wedge {\cal J}_u \wedge {\cal J}_v \wedge {\cal J}_r \wedge {\cal J}_s \;,
\eeq
we then have that,
\beq \label{d0rs}
d^{\hat{X}}_{\hat{1}rs} = \sum_{\Lambda<\Sigma} c_{\Lambda}^u c_{\Sigma}^v \tilde{d}_{uvrs} \;.
\eeq
This is a result that will be of use in what follows.

The intersection numbers $d^{\hat{X}}_{rst}$ enjoy a simpler relationship to the analgous quantities on $X$. Starting with equations \eqref{unsplitd1} and \eqref{unsplitmu} we find that we can rewrite the intersection numbers on $X$, using equation \eqref{dtilde}, as follows.
\beq \label{dflat}
d^{X}_{rst} = \int_{{\cal A}} c^u {\cal J}_u \wedge \mu^{{\cal C}} \wedge {\cal J}_r \wedge {\cal J}_s \wedge {\cal J}_t = \tilde{d}_{urst} c^u
\eeq
However, using equations \eqref{splitd1} and \eqref{splitmu}, together with \eqref{dtilde} we find that,
\bea
d^{\hat{X}}_{rst} &=& \int_{{\cal A} \times \mathbb{P}^n} \mu^{{\cal C}} \wedge(\wedge_{\Lambda=1}^{n+1} ({\cal J}_{\hat{1}}+ c_{\Lambda}^u {\cal J}_u)) \wedge {\cal J}_r \wedge {\cal J}_s \wedge {\cal J}_t  \\
&=& \int_{{\cal A}} \mu^{{\cal C}} \wedge ( \sum_{\Lambda} c_{\Lambda}^u {\cal J}_u) \wedge {\cal J}_r \wedge {\cal J}_s \wedge {\cal J}_t \\
&=& c^u \int_{{\cal A}}  \mu^{{\cal C}}  \wedge {\cal J}_u \wedge {\cal J}_r \wedge {\cal J}_s \wedge {\cal J}_t \\
&=& \tilde{d}_{urst}c^u \label{dsharp}
\eea
Comparing equations \eqref{dflat} and \eqref{dsharp} we find the simple relation.
\beq \label{drel}
d^{\hat{X}}_{rst} = d^{X}_{rst}
\eeq

\subsection{Second Chern Classes}

The second Chern class of a complete intersection manifold, $M$, with configuration matrix $ \left[ {\cal A}' | {\cal M}' \right]$ is given by the following expression.
\beq \label{generalc2}
c_2(TM) = c_2(TM)^{IJ} J_I \wedge J_J = \left[ \frac{1}{2}(- \delta^{IJ}(n_I+1) + \sum_A {\cal M}'^I_A {\cal M}'^J_A) \right] J_I \wedge J_J 
\eeq
Here $I,J$ run over the projective space factors in the ambient space ${\cal A}'$ and $A$ runs over the defining polynomials.

Applying this to the specific case of the split configuration matrix of $\hat{X}$ in \eqref{gen_split_app}, we find the following expression.
\beq\label{vecdefs}
c_2(T\hat{X})^{RS}=\frac{1}{2} \left[ - \delta^{RS}(n_R+1)+ \sum_{a=1}^{K-1} \left( \begin{array}{c} 0 \\ {\cal C} \end{array}\right)^R_a \left( \begin{array}{c} 0 \\ {\cal C} \end{array}\right)^S_a + \sum_{\Lambda=1}^{n+1} \left( \begin{array}{c}1  \\ {\bf c} \end{array} \right)^R_{\Lambda} \left( \begin{array}{c}1  \\ {\bf c} \end{array} \right)^S_{\Lambda} \right]
\eeq
Concentrating on some specific components we then find the following expressions.
\bea
c_2(T\hat{X})^{\hat{1} r}&=&\frac{1}{2} \sum_{\Lambda=1}^{n+1} c_{\Lambda}^r = \frac{1}{2} c^r \label{mr1}  \\
c_2(T\hat{X})^{r s}&=& \frac{1}{2} \left[- \delta^{rs}(n_r+1) + \sum_{a=1}^{K-1} {\cal C}_a^r{\cal C}_a^s + \sum_{\Lambda=1}^{n+1} c_{\Lambda}^r c_{\Lambda}^s \right] \\
&=& \frac{1}{2} \left[- \delta^{rs}(n_r+1) +  \sum_{a=1}^{K-1} {\cal C}_a^r{\cal C}_a^s + c^r c^s \right] - \sum_{\Lambda< \Sigma} c_{\Lambda}^{(r}c_{\Sigma}^{s)} \\
&=& c_2(TX)^{rs} - \sum_{\Lambda< \Sigma} c_{\Lambda}^{(r}c_{\Sigma}^{s)} \label{mr2} \\
c_2(T\hat{X})^{\hat{1}\hat{1}} &=& \frac{1}{2} \left[ -(n+1)+ \sum_{\Lambda}^{n+1} 1 \right] =0 \label{mr3}
\eea
Thus far these expression do not seem to give any particularly simple relations between the second Chern classes of $X$ and $\hat{X}$. We can obtain such a simple relation, however, by considering the contraction of the above quantities with the triple intersection numbers.
\bea
c_2(T\hat{X})_r &:=& d^{\hat{X}}_{rST} c_2(T\hat{X})^{ST}=d^{\hat{X}}_{rst} c_2(T\hat{X})^{st}+2 d^{\hat{X}}_{r s\hat{1}} c_2(T{\hat{X}})^{s\hat{1}} \\
&=& d^{X}_{rst} (c_2(TX)^{st} - \sum_{\Lambda<\Sigma} c_{\Lambda}^r c_{\Sigma}^s) + 2 \sum_{\Lambda <\Sigma} c_{\alpha}^u c_{\Sigma}^v \tilde{d}_{uvrs} \frac{1}{2} c^s \\
&=&c_2(TX)_r
\eea
In the above we have used the expressions in equations \eqref{mr1}, \eqref{mr2} and \eqref{mr3}, together with the expressions involving intersection numbers given in equations \eqref{drel} and \eqref{d0rs}, and finally the definition given in \eqref{dtilde}.
\section{Bundle-valued Cohomology on CICYs}\label{coh_appendix}
The main results of this paper rely heavily on computations of vector bundle-valued cohomology on Complete Intersection Calabi-Yau three-folds. As a result, it is worth reviewing here a few general results and a collection of useful tools for explicit computations of line bundle-valued cohomology. More complete treatments of these tools and techniques are available in \cite{hubsch,Anderson:2008ex}.

For ease of computation, the examples in this paper have all been built from line bundles, $\cL$ on $X$ which are ``favorable" in the sense defined in Section \ref{vac_var_sec}. That is, they descend from line bundles on an ambient product of projective spaces. As a result, we will focus on a series of techniques to compute line bundle cohomology on $X$ using information, structure and maps from the ambient space ${\cal A}$. It is important to note however, that complex-structure fixing vector bundles of the form \eref{ext1}\footnote{More generally, those bundles holomorphic only on higher co-dimensional loci in the complex structure moduli space of $X$.} {\it do not descend from the ambient space}. That is, for the extension bundles $0 \to \cL \to V \to \cL^{\vee} \to 0$, while $\cL, \cL^{\vee}$ are favorable, a non-trivial extension such as $V$ can be defined only on $X$ (and only for specific values of the complex structure). With these distinctions in mind, we turn now to techniques for determining bundle cohomology for favorable bundles descending from ${\cal A}$.
 
\subsection{The Koszul Resolution}\label{koszul_rev}
The standard method of computing the cohomology of a vector bundle
$V= \cV|_X$  coming from the restriction of $\cV$ from an ambient space $\cA$ to
the variety $X$ is the so-called {\it Koszul Resolution} of
$V|_X$. In general, if $X$ is a smooth hypersurface of co-dimension $K$, which is the zero locus of a holomorphic section $s$ of the bundle $N$, then the following long exact sequence exists \cite{AG}:
\beq\label{koszulB}
0 \to \cV \otimes \wedge^K \cN^{\vee} \to \cV \otimes \wedge^{K-1} \cN^{\vee}
\to \ldots \to \cV \otimes \cN^{\vee} \to \cV \to \cV|_X \to 0 \ .
\eeq
where the fiber-wise morphisms appearing in the sequence arise from appropriate sections, $H^0({\cal A}, \wedge^j \cN)$. If the cohomology of the bundles $\wedge^j \cN^{\vee} \otimes \cV$ are known on the ambient space, we can use the Koszul sequence to determine the cohomology of $V|_X$. Here, $\cN^{\vee}$ is the dual to the normal bundle. We recall that for a CICY, the normal bundle to the space is given by the configuration matrix \eref{config}:
\beq\label{normal2}
\cN = \bigoplus_{j=1}^K \cO(q_j^1, \ldots, q_j^m) \ .
\eeq
In the above, we have generalized the standard notation that
$\cO_{\IP^n}(k)$ denotes the line-bundle over $\IP^n$ 
whose sections are degree $k$ polynomials in the coordinates of $\IP^n$; that is, $\cO(q^j_1,
\ldots, q^j_m)$ is the line-bundle over $\IP^{n_1} \times \ldots
\times \IP^{n_m}$ whose sections are polynomials of degree $q^j_1,
\ldots, q^j_m$ in the respective $\IP^{n_i}$-factors. Being a direct
sum, the rank of $\cN$ is $K$. 

We can break the sequence \eref{koszulB} into a series of short exact sequences by introducing appropriate cokernels, ${\cal K}_i$, as 
\bea
0 \to \cV \otimes \wedge^K \cN^{\vee} \to \cV \otimes \wedge^{K-1} \cN^{\vee}
\to \mathcal{K}_1 \to 0 \\
0 \to \mathcal{K}_1 \to \cV \otimes \wedge^{K-2} \cN^{\vee} \to \mathcal{K}_2 \to 0 \\
\ldots \\
0 \to \mathcal{K}_{K-1} \to \cV \to \cV|_X \to 0
\eea
and each of these short exact sequences will give rise to a long exact sequence in cohomology:
\bea \label{long_exact}
0&\to& H^0(\cA, \cV \otimes \wedge^K \cN^{\vee}) \to H^0(\cA, \cV \otimes \wedge^{K-1} \cN^{\vee})
\to H^0(\cA, \mathcal{K}_1) \\
0 &\to& H^0(\cA,\mathcal{K}_1) \to H^0(\cA, \cV \otimes \wedge^{K-2} \cN^{\vee}) \to H^0(\cA, \mathcal{K}_2) \to \ldots \\
\ldots \\
0 &\to& H^0(\cA, \mathcal{K}_{K-1}) \to H^0(\cA, \cV) \to H^0(X, \cV|_X) \to \ldots
\eea
To find $H^*(X, \cV|_{X})$ we must determine the various cohomology groups in \eref{long_exact}. It is easy to see that for higher co-dimensional spaces or tensor powers of bundles, this decomposition of sequences is a laborious process. Fortunately, the analysis of these arrays of exact sequences is dramatically simplified by the use of spectral sequences. Spectral sequences \cite{AG} are completely equivalent to the collection of exact sequences described above, but can be useful for such lengthy cohomology computations. 

\subsection{The Spectral Sequence}\label{s:leray}
To obtain the necessary cohomology of $\cV|_X$ from \eref{koszulB}, we define a tableaux
\beq\label{leray}
E^{j,k}_{1}(V) := H^j({\cal A},  \cV \otimes \wedge^{k} \cN^{\vee}), \qquad
k = 0, \ldots, K; \  j=0, \ldots, \dim(A) = \sum_{i=1}^m n_i \ .
\eeq
This forms the first term of a so-called {\it Koszul spectral sequence} \cite{AG, Green:1987cr}. The spectral sequence is a complex defined by differential maps $d_i : E^{j,k}_i \to
E^{j-i+1,k-i}_i$ for $j = 1,2,\ldots$ {\it ad infinitum} where $d_i \circ d_i =0$. The subsequent terms in the spectral sequence are defined by
\beq \label{leray_iterate}
E^{j,k}_{i+1}(V)= \frac{ker(d_{i}: E^{j,k}_{i}(\cV) \rightarrow E^{j-i+1,k-i}_{i}(\cV))}{Im(d_{i}: E^{j+i-1,k+i}_{i}(\cV) \rightarrow E^{j,k}_{i}(\cV))}
\eeq
Since the number of terms in the Koszul sequence  \eref{koszulB} is finite, there exists a limit to the spectral sequence. That is, the sequence of tableaux converge after a finite number of steps to $E^{j,k}_{\infty}(\cV)$. The actual cohomology of the bundle $V=\cV|_{X}$ is constructed from this limit tableaux:
\beq\label{hodge_converge}
h^q(X, V) = \sum^{K}_{m=0} \text{rank} E^{q+m,m}_{\infty}(\cV) \ .
\eeq
where $h^q(X, V)= \text{dim} (H^q(X, V))$. 

In practice, the tableaux $E^{p,q}_r$ converges fairly rapidly because many of its entries will turn out to be zero and the associated maps $d_i$, vanish; hence the spectral sequence converges after only a few steps. However, in general all computations which involve long exact cohomology sequences \eref{long_exact} or associated spectral sequences \eref{leray_iterate} rely upon the ability to discern the action of maps between cohomology groups on the ambient space $\mathcal{A}$. In fortunate cases, the tableau are sufficiently sparse that is possible to determine the required dimensions of cohomology groups without knowing any maps explicitly. But in general the obstacle cannot be avoided. Fortunately, the task of computing the rank and kernels of the spectral maps can be accomplished straightforwardly for favorable bundles on CICYs using the coset representation of Flag spaces and the tensor algebra associated with representations of Lie groups \cite{hubsch}. 

\subsection{Cohomology of line bundles on CICYs}
Up to this point, our comments on bundle cohomology has been general. However, we focus  now on the particular case of most interest to us in this work: Line Bundle Cohomology on CICYs. The first important tool in our arsenal is a computational variation on the Bott-Borel-Weil theorem \cite{barton, hubsch}.

\subsubsection{Flag Spaces and the Bott-Borel-Weil Theorem}\label{bbw}

It can be shown that every simply connected compact homogeneous complex space is homeomorphic to a torus-bundle over a product of certain coset spaces $G/H$, where $G$ is a compact simple Lie group and $H$ is a regular semi-simple subgroup. Such spaces are known as C-spaces or `generalized flag varieties' \cite{hubsch}. In fact, the simplest example of this is $\mathbb{P}^{n}= (\frac{U(n+1)}{U(1) \times U(n)})$. Viewing $\mathbb{P}^n$ in this way will prove useful to us since it can be shown that homogeneous holomorphic vector bundles over such flag varieties, $\mathbb{F} = (G_{\mathbb(C)}/H)$, are labeled by representations of $H$ (for our applications, $H=U(1) \times U(n)$). This will provide us with a powerful new tool to investigate bundle cohomology on CICYs.

Recalling that a representation can be written as a direct sum of irreducible ones, we can focus on irreducible homogeneous holomorphic vector bundles. Further, we know that such representations are uniquely labeled by their highest weight, so we have a convenient notation for such bundles. For this purpose, we will use the well-known Young tableaux (see e.g. \cite{Georgi:1982jb}). We will be dealing strictly with unitary groups and will adopt the following conventions. To denote a bundle, we write $(a_{1}, \ldots, a_{n})$, where $a_r \leq a_{r+1}$ is the number of boxes in the rth row of the tableau. For $a_r >0$ ($<0$) the boxes are arrayed to the right (left) of the 'spine'. Therefore, in the standard tensorial notation, $(-1,0, \ldots 0)$ denotes a covariant vector $v_{\mu}$ while $(0, \ldots ,0,1)$ labels the contravariant vector $v^{\mu}$. All other representations can be obtained from these by multiplication and then decomposition into a direct sum of irreducible components through symmetrizing, anti-symmetrizing and taking traces with the invariant tensor (${\delta}^{\mu}_{\nu}$). A tensor product of representations of factor $U(n_{f})$'s can be written as the Young tableau,
\beq
(a_{1}, \ldots a_{n_1} | b_{1}, \ldots , b_{n_2} | \dots |d_{1}, \ldots , d_{n_F})
\eeq
or for a more condensed notation, we can stack the partitions vertically on top of each other. 

For the case of line bundles, we recall that we may view $\mathbb{P}^n$ as the space of all lines $L \approx \mathbb{C}^1$ through the origin of $\mathbb{C}^{n+1}$. Each line is defined as the zero set of some linear polynomial $l(x)$ over $\mathbb{C}^{n+1}$. Now, from the definition of the hyperplane bundle $\cO(1)$ on $\mathbb{P}^n$ as the line bundle whose (global holomorphic) sections are linear polynomials we may formulate a line bundle in the language of flag spaces above. Viewing $\mathbb{P}^n$ as a quotient of unitary groups and a bundle over it as a representation of $U(1) \times U(n)$, a little thought reveals that we may denote $\cO(1)$ as $(-1|0, \ldots 0)$ (and similarly, its dual bundle $\cO(-1)$ is written $(1|0, \ldots 0)$). 

With this notation in hand, let $\mathbb{F}= \frac{U(N)}{U(n_1) \times \ldots \times U(n_F)}$ (with $N = \sum_{f} n_{f}$) be a flag space as above and $V$ be a holomorphic homogeneous vector bundle over it. Then
\begin{theorem}{Bott-Borel-Weil} \label{BBW}

(1) Homogeneous vector bundles $V$ over $\mathbb{F}$ are in 1-1 correspondence with the $U(n_1) \times ...\times U(n_F)$ representations.

(2) The cohomology $H^{i}(\mathcal A,V)$ is non-zero for at most one value of $i$, in which case it provides an irreducible representation of $U(N)$, $H^{i}(\mathbb{F},V) \approx (c_1,...,c_N)\mathcal{C}^N$.

(3) The bundle, $(a_1,...,a_{n_1}|...|b_1,...,b_{n_F})$, determines the cohomology group $(c_1,...,c_N)$, according to the following algorithm:

1. Add the sequence $1..., N$ to the entries in $(a_1,...,a_{n_1}|...|b_1,...,b_{n_F}) $.

2. If any two entries in the result of step 1 are equal, all cohomology vanishes; otherwise proceed.

3. swap the minimum number ($=i$) of neighboring entries required to produce a strictly increasing sequence.

4. Subtract the sequence $1,...N$ from the result of $3$, to obtain $(c_{1},c_{2},...,c_{N})$.
\end{theorem}

Using this algorithm, it is straightforward to reproduce the Bott-formula \cite{AG,hubsch} for cohomology of line bundles on single projective spaces by simply counting the dimensions of the the associated Young tableau $(c_{1},c_{2},...,c_{N})$ of the unitary representations. The result is 

\beq\label{bottformula}
h^{q}(\mathbb{P}^{n}, \mathcal{O}_{\IP^n}(k))=\left\{
\begin{array}
[c]{ll}%
\binom{k+n}{n} & q=0\quad k>-1\\
1 & q=n\quad k=-n-1\\
\binom{-k-1}{-k-n-1} & q=n\quad k<-n-1\\
0 & \mbox{otherwise}
\end{array}
\right. \ .
\eeq

where the binomial coefficients arise from the dimensions of Young tableau (see \cite{Georgi:1982jb} for a review of the hook-length formulas). 

The computation of line bundle cohomology described by the Bott-Borel-Weil theorem is easily generalized to products of projective space using the K\"unneth formula \cite{AG} which gives the cohomology of bundles over a direct product of
spaces. For products of projective spaces it states that:
\beq\label{kunneth}
H^n(\IP^{n_1} \times \ldots \times \IP^{n_m}, \cO(q_1, \ldots, q_m)) =
\bigoplus_{k_1+\ldots+k_m = n} H^{k_1}(\IP^{n_1},\cO(q_i)) \times
\ldots \times H^{k_m}(\IP^{n_m},\cO(q_m)) \ ,
\eeq

With this in hand, we can compute the cohomology of line bundles over the ambient space. For example, in the notation of flag varieties, the line bundle $l = \mathcal{O}(k_{1},-k_{2})$ on $\mathbb{P}^{n_1} \times \mathbb{P}^{n_2}$ (with $k_{2} \geq n_{2}+1$) can be denoted by a product of irreps of $(U(1) \times U(n_{1})) \times (U(1) \times U(n_{2}))$:
\beq
l \sim \binom{-k_{1}|0,\ldots ,0}{{~k_2}|0, \ldots, 0}
\eeq
where there are $n_1$ zeroes in the first row and $n_2$ zeroes in the second. Using Bott-Borel-Weil and the Kunneth formula then, the cohomology of this line bundle on the ambient space would be described by 
\beq \label{sample_coho}
H^{n_{2}}(\cA , l) \sim \binom{-k_{1},0,\ldots ,0}{1, \ldots, 1,({k_2}-n_{2})}
\eeq
where $(-k_{1},0,\ldots ,0)$ denotes the Young Tableau of a irreducible representation of $U(n_{1} +1)$,  $(1, \ldots, 1,({k_2}-n_{2}))$ is the Young tableau of a $U(n_{2}+1)$ irrep and the Kunneth product of the restricted cohomology groups is denoted by the vertical stacking of tableau. We recall that the dimension of a Young tableau may be easily computed from the hook-length formula (see \cite{Georgi:1982jb}, for example). For instance, the dimension of $(-k_{1},0,\ldots ,0)$ is just the degrees of freedom in a totally symmetric tensor in $(n_{1}+1)$ variables, namely $\binom{k_{1} + n_{1}}{n_{1}} $. In counting the degrees of freedom in the tableau $(1, \ldots, 1,({k_2}-n_{2}))$, it is useful to recall that the totally anti-symmetric tensor, $\epsilon^{[a,...,b]}$ is a singlet under $U(n)$. Thus we can strip a Levi-Civita tensor from the tableau $(1, \ldots, 1,({k_2}-n_{2}))= (1, \ldots, 1) \otimes (0, \ldots , 0,({k_2}-n_{2}-1))$ and just consider the dimension of $(0, \ldots , 0,k{_2}-n_{2}-1)$ which is yet another symmetrized tensor whose degrees of freedom may be counted as before. Therefore, the total cohomology/tableau $\binom{-k_{1},0,\ldots ,0}{1, \ldots, 1,({k_2}-n_{2})}$ has dimension $\binom{k_{1} + n_{1}}{n_{1}} \times \binom{k_{2}-1}{n_{2}}$. 

In summary, by using the Bott-Borel-Weil theorem we are able to represent the cohomology groups of line bundles over the ambient space, $\mathcal{A}$, as irreducible representations of unitary groups (and readily compute their dimensions). Returning to the task of computing the line bundle cohomology on the Calabi-Yau $3$-fold, $X$, we note that this technique will dramatically simplify the spectral sequence calculations of the previous section by  providing a simple representation for the ambient space cohomology groups involved. We will reduce the abstract task of determining the properties of maps between line bundle cohomology groups to the more straightforward one of studying maps between irreps of unitary groups. 

\subsubsection{Applying Bott-Borel-Weil}
Since the previous section was somewhat abstract, here we will illustrate and apply the results of the Bott-Borel-Weil theorem in a simple way and summarize the main tools that we need for this paper. From the previous section, it is clear that cohomology groups on an ambient space of the form ${\cal A}=\mathbb{P}^{n_1}\times \ldots \mathbb{P}^{n_m}$, can be represent by irreps of unitary groups -- i.e. fully symmetrized tensors. More specifically, let us summarize the necessary ingredients. First, for a single projective space $\mathbb{P}^n$ consider the cohomology of the line bundle $\cO(k)$ with $k>0$. From the Bott theorem and the algorithm described in Theorem \ref{BBW}, the only non-vanishing cohomology group is $H^0(\mathbb{P}^n, \cO(k))$ and the elements of this group can be represented by a fully symmetrized tensor with
$k$-indices. We will choose the convention that for $k>0$ those indices are ``down" type:
\beq
H^0(\mathbb{P}^n, \cO(k)) \leftrightarrow f_{(a_1 \ldots a_k)}
\eeq
As might be expected since $H^0(\mathbb{P}^n, \cO(k))$ is the group of global sections of $\cO(k)$, this cohomology group is also space of polynomial functions over ${\cal A}$ of degree $k$. Moreover, this is compatible with the tensor description above since we can simply view the tensor $f_{(a_1, \ldots, a_k)}$ as the coefficients of a general polynomial $p \in H^0(\mathbb{P}^n, \cO(k))$. That is,
\beq
p=f_{(a_1 \ldots a_k)}x^{a_1}x^{a_2}\ldots x^{a_k}
\eeq
where $x^a$, $a=1,\ldots (n+1)$ runs over the homogeneous coordinate on $\mathbb{P}^n.$ Likewise from \eref{bottformula}, for $\cO(-k)$, the only non-vanishing cohomology is $H^n(\mathbb{P}^{n}, \cO(-k))$ and it can be represented by the product of the unique fully antisymmetric tensor in $n+1$ indices and a fully symmetrized tensor with $k-(n+1)$ indices, both of ``up" type:
\beq
H^n(\mathbb{P}^{n}, \cO(-k)) \leftrightarrow \epsilon^{[a_1 \ldots a_n]}g^{(b_1 \ldots b_{(k-(n+1))})}
\eeq
Stripping off the antisymmetric tensor $\epsilon$, we can represent the tensor $g^{(b_1 \ldots b_{(k-(n+1))})}$ in a similar way in terms of a ``polynomial" space. This time, though involving either ``inverse" polynomials or ``derivatives". For example, $q \in H^n(\mathbb{P}^n, \cO(-k))$ could be represented
\beq\label{poldef}
q=g^{(b_1 \ldots b_{(k-(n+1))})}\frac{1}{x^{b_1}}\ldots\frac{1}{x^{(k-(n+1))}}
\eeq
Or equivalently
\beq\label{derdef}
q=g^{(b_1 \ldots b_{(k-(n+1))})}\partial x_{b_1}\ldots \partial x_{(k-(n+1))}
\eeq
In either case, the every cup product in cohomology (i.e Yoneda pairing) can be represented by polynomial operations. For instance, in terms of the Bott-Borel-Weil tensor representations, the following product
\beq\label{samplecup}
H^0(\mathbb{P}^n, \cO(k)) \wedge H^n(\mathbb{P}^n, \cO(-(k+n+1))) \Rightarrow H^n(\mathbb{P}^n, \cO(-(n+1))) \simeq \mathbb{C}
\eeq
would be described in terms of tensor contraction (suppressing irrelevant epsilon tensors on both sides of the expression) as
\beq\label{contrac}
f_{(a_1 \ldots a_k)}g^{(a_1 \ldots a_{k})} \rightarrow \mathbb{C}
\eeq
This same contraction can be accomplished in a polynomial representation by describing $H^n(\mathbb{P}^n, \cO(-(k+n+1)))$ via ``inverse" polynomials, where the multiplication rule takes the form
\beq
x^a \left( \frac{1}{x^{b}} \right)= \delta^{a}_{b}
\eeq
Thus, the map in \eref{samplecup} is schematically
$(\text{poly. of deg(k)}) \left (\frac{1}{\text{poly. of deg(k)}} \right) \Rightarrow \mathbb{C}$:
\beq
f_{a_1 \ldots a_k}g^{b_1 \ldots b_{k}}\left( x^{a_{1}\ldots a_{k}}\frac{1}{x^{b_1}}\ldots\frac{1}{x^{b_k}} \right) \Rightarrow \mathbb{C}
\eeq
Similarly, $H^n(\mathbb{P}^n, \cO(-(k+n+1)))$ can be represented by derivatives with the obvious ``multiplication rule"
\beq
\partial x_a (x^b) =\delta^{b}_{a}
\eeq
and \eref{contrac} can be calculated via
\beq
f_{a_1 \ldots a_k}g^{b_1 \ldots b_{k}}\partial x_{b_1} \ldots \partial x_{b_k}\left(  x^{a_{1}\ldots a_{k}} \right) \Rightarrow \mathbb{C}
\eeq
So long as care is taken with relative constant prefactors, either of the two descriptions in \eref{poldef} and \eref{derdef} can be used to represent cohomology groups of the form $H^n(\mathbb{P}^n, \cO(-k))$.

The generalization to products of projective spaces is immediate: For a cohomology group of the form $H^0({\cal A}, \cO(k_1,\ldots k_m))$ with $k_i >0$ for $i=1,\ldots m$ for example, we have a tensor
\beq
f_{(a_1, \ldots, a_{k_1})(b_1 \ldots b_{k_2})\ldots (c_{1}\ldots c_{k_m})}
\eeq
where each index type ranges over the appropriate range for the given $\mathbb{P}^{n_{i}}$ factor. Furthermore, ``mixed" tensors, with both ``up" and ``down"-type indices, now arise. For example in ${\cal A}=\mathbb{P}^{2}\times \mathbb{P}^{3}$, we could represent $H^2({\cal A}, \cO(-4,2))$ via tensors as
\beq
\epsilon^{[abc]}{f^{d}}_{(\alpha\beta)}
\eeq
where $a=0,1,2$ labels the homogeneous coordinates, $x^a$, of $\mathbb{P}^2$ and $\alpha=0,1,2,3$ counts the homogeneous coordinates, $y^{\alpha}$, of $\mathbb{P}^3$. 

The polynomial formalism described above applies immediately to the ``multi-degree" polynomials of a product of projective spaces. As final example, consider the following map in ambient space cohomology
\beq
\phi: H^2({\cal A}, \cO(-4,2)) \to H^2({\cal A}, \cO(-3,3))
\eeq
Suitable polynomial representatives take the form (again, stripping $\epsilon^{[abc]}$ from both the source and target of the map):
\begin{align}
&H^2({\cal A}, \cO(-4,2)) :& {f^{d}}_{(\alpha\beta)} \frac{1}{x^d}y^{\alpha}y^{\beta} \\
&H^2({\cal A}, \cO(-3,3)) :&g_{(\rho \sigma \delta)} y^{\rho}y^{\sigma}y^{\delta} \\
&\phi \in H^0({\cal A}, \cO(1,1)) : & \phi_{a \alpha} x^a y^\alpha
\end{align}
As demonstrated in the main body of the text, such polynomial representatives make it possible to explicitly compute the ranks of all the maps in cohomology considered in this work.

\section{Proof of Lemma 1}\label{app_proof}
In this section, Lemma \ref{lemma1} is proved for a general $\mathbb{P}^n$ split of the form \eref{gen_split}. To begin, let us restate the result here:
\begin{UnLemma} \label{lemma_again}
Suppose that $X$ and ${\hat X}$ are two Calabi-Yau three-folds realized as complete intersections in products of projective spaces, related by a ``splitting transition" of the type described in \eref{gen_split}. Let $\cL=\cO(a,\ldots, b)$ be a ``favorable" line bundle on $X$ -- that is, a line bundle corresponding to a divisor $D \subset X$ such that $D={\cal D}|_X$ is the restriction of a divisor, ${\cal D}$, in a $\mathbb{P}^{n_i}$ factor of the ambient space. Then the calculation (and dimension) of the cohomology of ${\hat \cL}=\cO(0\ldots ,0,a,\ldots, b)$ is identical to that of $L$ on the ``Derminantal locus" (defined by \eref{gen_split} and \eref{detvardef}) shared in the complex structure moduli space of $X$ and ${\hat X}$.
\end{UnLemma}

While this lemma holds for arbitrary cohomology of $\cL$ on $X$, for the sake of explicitness, we will provide the proof here for $H^1(X,\cL)$, the cohomology group defining non-trivial Extensions of line bundles used throughout this work.

To begin, consider the Koszul sequence associated to the line bundle $L$ on $X$
\bea\label{kozulapp}
0 \to \cL_{\cA} \otimes \wedge^K \cN^{\vee} \to \cL_{\cA} \otimes \wedge^{K-1} \cN^{\vee}
\to \mathcal{K}_1 \to 0 \\
0 \to \mathcal{K}_1 \to \cL_{\cA} \otimes \wedge^{K-2} \cN^{\vee} \to \mathcal{K}_2 \to 0 \\
\ldots \\
0 \to \mathcal{K}_{K-1} \to \cL_{\cA} \to \cL \to 0
\eea

Without loss of generality, we will assume here that $H^1({\cal A}, \cL_{\cA})=0$ (recall that $H^i({\cal A}, \cL_{\cA}) \neq 0$ for at most one value of $i$). Then in general, the long exact sequences in cohomology associated to \eref{kozulapp} give $H^1(X,\cL)$ as
\beq
H^1(X,\cL)= \text{ker}(\phi),~~~\phi: H^2(\cA, \mathcal{K}_{K-1}) \to H^2(\cA, \cL_{\cA})
\eeq
where $H^2(\cA, \mathcal{K}_{K-1})$ could have contributions from each of
\beq
H^{j+1}(\cA, \wedge^{j} \cN^{\vee}\otimes \cL_{\cA})
\eeq 
for $j=1, \ldots K$. Again, for succinctness, we will for the moment assume that only one of these cohomology groups is non-zero (note that in the case in which multiple cohomology groups are non-vanishing, the map arguments below can simply be repeated for each map individually). Then,
\begin{align}\label{thegoodphi}
&H^1(X,\cL)= \text{ker}(\phi),~~~\phi: H^2(\cA, \mathcal{K}_{K-1}) \to H^2(\cA, \cL_{\cA}) \\
&H^2(\cA, \mathcal{K}_{K-1}) \simeq H^{j'+1}(\cA, \wedge^{j'} \cN^{\vee}\otimes \cL_{\cA})
\end{align}
for some $j'$. Using the techniques of the previous section we can represent this polynomial/inverse-polynomial multiplication where the map in question, $\phi$, is a global section of $H^0(\cA,\wedge^{j'} \cN^{\vee})$.

Now, consider a general ``splitting" of $X$,
\beq\label{gen_split_app_again}
X=\left[ {\cal A} |{\bf c}\;  {\cal C} \right] \longrightarrow {\hat X}= \left[ \begin{array}{c|c c c c c}
{\mathbb P}^n & 1 & 1 & \ldots & 1 & {\bf 0} \\
{\cal A} & {\bf c}_1 & {\bf c}_2 & \ldots & \bf{c}_{n+1} & {\cal C}
\end{array} \right]
\eeq
 which defines a new manifold ${\hat X}$. We will study the cohomology of ${\hat \cL}=\cO(0,\ldots, 0, a,\ldots, b)$ on that space. 
 
 As in Appendix \ref{split_idens}, and equation \eref{vecdefs}, we note that the normal bundle of ${\hat X}$ takes the schematic form
 \beq\label{norm_breakdown}
 \cN_{\hat X} \simeq  \bigoplus \cN_{split}\oplus \cN_{{\cal C}}  \simeq   \left( \begin{array}{c}1  \\ {\bf c} \end{array} \right) \oplus \left( \begin{array}{c} 0 \\ {\cal C} \end{array}\right)
 \eeq
where the first term consists of $K-1$ line bundles and the second term is the $n+1$ line bundles arising from the splitting of a column of the configuration matrix of $X$.

Now, for the line bundle ${\hat \cL}$ as defined above, we must consider how the computation of $H^1({\hat X}, {\hat \cL})$ compares to that of $X$ above. First, by the definition of ${\hat \cL}$ and the Bott-Formula, \eref{bottformula}, we note the following isomorphism
\beq
H^2({\hat \cA}, \hat{\cL}_{{\hat \cA}}) \simeq H^2(\cA, \cL_{\cA})
\eeq
Likewise, if we look in detail at $H^{j'+1}(\cA, \wedge^{j'} \cN^{\vee}\otimes \cL)$, we will see that the structure of the non-vanishing cohomology groups is determined by the form of the normal bundle in \eref{norm_breakdown}. Any contributions to non-trivial cohomology from  $\cN_{{\cal C}}$ will carry through into $H^{j'+1}(\hat{\cA},  \wedge^{j'} \cN_{\hat{X}}^{\vee}\otimes {\hat \cL}_{{\hat \cA}})$ (by the zero entries in the new $\mathbb{P}^n$ direction, the Kunneth Formula, \eref{kunneth} and \eref{bottformula}). Thus, the only new non-trivial contributions to the cohomology of ${\hat \cL}$ must arise from powers of $\wedge^{k} \cN_{\hat X}$ involving $\cN_{split}$.

For these, we note that regardless of the degree of the ${\bf c}$ entries, since each component of $\cN_{split}$ in the new $\mathbb{P}^n$ directions is a ``1", when we consider $(\wedge^k \cN^{\vee}_{split} \otimes {\hat L}_{{\hat \cA}})$ the Bott-Formula, \eref{bottformula}, for the $\mathbb{P}^n$ direction guarantees that, in fact, only one cohomology group can possibly be non-trivial, namely
$H^{j'+1+n}({\hat A}, \wedge^{j'+n}\cN^{\vee}_{split} \otimes {\hat \cL}_{{\hat \cA}})$. Thus, for ${\hat \cL}$ on ${\hat X}$ the cohomology map of interest (i.e. the equivalent of \eref{thegoodphi}) takes the form
\begin{align}\label{phihatmap}
&{\hat \phi}: H^2({\hat \cA}, {\hat {\mathcal K}}_{K'-1}) \to H^2({\hat \cA}, {\hat \cL}_{{\hat \cA}}) \\
&H^2({\hat \cA}, {\hat {\mathcal K}}_{K'-1}) \simeq H^{j'+1+n}({\hat A}, \wedge^{j'+n}\cN^{\vee}_{split} \otimes {\hat \cL}_{{\hat \cA}}) 
\end{align}
Now finally, we can compare $H^{j'+1+n}({\hat A}, \wedge^{j'+n}\cN^{\vee}_{split} \otimes {\hat \cL}_{{\hat \cA}})$ and $H^{j'+1}(\cA, \wedge^{j'} \cN^{\vee}\otimes \cL_{\cA})$. Again, using the zero-entries in the new $\mathbb{P}^n$ direction as well as the Bott and Kunneth Formulae, we find
\beq
H^{j'+1+n}({\hat A}, \wedge^{j'+n}\cN^{\vee}_{split} \otimes {\hat \cL}_{{\hat \cA}})=H^n(\mathbb{P}^n, \cO(-n-1)) \times H^{j'+1}(\cA, \wedge^{j'} \cN^{\vee}\otimes \cL_{\cA})
\eeq
Thus, using the tensor/polynomial descriptions of cohomology from the previous section, we see that the only difference between the source and target in \eref{thegoodphi} and \eref{phihatmap} is a factor of
\beq
H^n(\mathbb{P}^n, \cO(-n-1)) \simeq \epsilon^{(a_1 \ldots a_{n+1})}
\eeq
in the left hand side (i.e. source) of \eref{phihatmap}.

Meanwhile the maps in question are $\phi \in H^0(\cA, \wedge^{j'} \cN_{X})$ and ${\hat \phi} \in H^0({\hat \cA}, \wedge^{j'+n}\cN_{split})$. It is straightforward to verify that
\beq
\epsilon^{(a_1 \ldots a_{n+1})}{\hat \phi}^{1}_{a_{1}}\ldots {\hat \phi}^{n+1}_{a_{n+1}} \simeq \phi_{split}
\eeq
That is, the contraction on the left hand side defining map in cohomology for ${\hat X}$ is exactly the cohomology map, $\phi$ of $X$, tuned in complex structure moduli space to the ``Determinantal Variety" form of \eref{detvardef}. This establishes the above Lemma for the chosen map. Additional maps in cohomology follow in an entirely analogous manner and multiple splittings follow immediately by induction.

\end{document}